\documentclass[useAMS,usenatbib]{mn2e}
\usepackage{graphicx}
\usepackage{longtable}

\title[Physical parameters of the OGLE-LMC-CEP-0227]{Physical parameters and the projection factor of the classical Cepheid in the binary system OGLE-LMC-{\allowbreak}CEP-0227}
\author[Pilecki et al.]
{B. Pilecki$^{1,2}$\thanks{e-mail: pilecki@astrouw.edu.pl}, D. Graczyk$^{2}$\thanks{e-mail: darek@astro-udec.cl}, G. Pietrzy{\'n}ski$^{1,2}$, W. Gieren$^{2,11,12}$, I. B. Thompson$^{4}$,
\newauthor
 W. L. Freedman$^{4}$, V. Scowcroft$^{4}$, B. F. Madore$^{4}$, A. Udalski$^{1}$, I. Soszy{\'n}ski$^{1}$, 
\newauthor
 P. Konorski$^{1}$, R. Smolec$^{3}$, N. Nardetto$^{5}$, G. Bono$^{6,7}$, P. G. Prada Moroni$^{8,9}$,
\newauthor
 J. Storm$^{10}$, A. Gallenne$^{2}$ \\
$^{1}$Warsaw University Observatory, Al. Ujazdowskie 4, PL-00-478, Warszawa, Poland\\
$^{2}$Universidad de Concepci{\'o}n, Departamento de Astronom{\'i}a,Casilla 160-C, Concepci{\'o}n, Chile\\
$^{3}$Copernicus Astronomical Centre, Polish Academy of Sciences, Bartycka 18, 00-716 Warsaw, Poland\\
$^{4}$Carnegie Observatories, 813 Santa Barbara Street, Pasadena, CA 91101-1292, USA\\
$^{5}$Laboratoire Lagrange, UMR7293, UNS/CNRS/OCA, 06300 Nice, France\\
$^{6}$Dipartimento di Fisica Universit`a di Roma Tor Vergata, viadella Ricerca Scientifica 1, 00133 Rome, Italy\\
$^{7}$INAF - Osservatorio Astronomico di Roma, Via Frascati 33, 00040 Monte Porzio Catone, Italy\\
$^{8}$Dipartimento di Fisica ‘E. Fermi’, Universit`a di Pisa, Largo B. Pontecorvo, 3, I-56127, Pisa, Italy\\
$^{9}$INFN, Sezione di Pisa, Largo B. Pontecorvo, 3, I-56127, Pisa, Italy\\
$^{10}$Leibniz-Institut f ̈ur Astrophysik Potsdam (AIP), An der Sternwarte 16, D-14482 Potsdam, Germany\\
$^{11}$University Observatory Munich, Scheinerstrasse 1, 81679 Munich, Germany\\
$^{12}$Max Planck Institute for Extraterrestrial Physics, Giessenbachstrasse, 85748 Garching, Germany}

\begin{document}

\date{Accepted 2013 August 10.
      Received 2013 July 29;
      in original form 2013 July 3}
\pubyear{2013}
\maketitle
\label{firstpage}

\begin{abstract}

A novel method of analysis of double-lined eclipsing binaries containing a radially pulsating star is presented. The combined pulsating-eclipsing light curve is built up from a purely eclipsing light curve grid created using an existing modeling tool. For every pulsation phase the instantaneous radius and surface brightness are taken into account, being calculated from the disentangled radial velocity curve of the pulsating star and from its out-of-eclipse pulsational light curve and the light ratio of the components, respectively. The best model is found using the Markov Chain Monte Carlo method.

The method is applied to the eclipsing binary Cepheid OGLE-LMC-CEP-0227 ($P_{puls}=3.80$ d, $P_{orb}=309$ d). We analyze a set of new spectroscopic and photometric observations for this binary, simultaneously fitting OGLE V-band, I-band and {\it Spitzer} 3.6 $\mu$m photometry. We derive a set of fundamental parameters of the system significantly improving the precision comparing to the previous results obtained by our group. The Cepheid mass and radius are $M_1=4.165 \pm 0.032 \,M_\odot$ and $R_1=34.92 \pm 0.34 \,R_\odot$, respectively.

For the first time a direct, geometrical and distance-independent determination of the Cepheid projection factor is presented. The value $p=$1.21 $\pm$ 0.03(stat.) $\pm$ 0.04(syst.) is consistent with theoretical expectations for a short period Cepheid and interferometric measurements for $\delta$ Cep. We also find a very high value of the optical limb darkening coefficients for the Cepheid component, in strong disagreement with theoretical predictions for static atmospheres at a given surface temperature and gravity.

\end{abstract}


\section{Introduction}  
\label{sect:intro}

Radially pulsating stars like Cepheids, RR Lyrae or Miras stars are important distance indicators in the local universe. The presence of such a variable star in an eclipsing binary system serves as a unique opportunity to derive fundamental astrophysical parameters of the pulsating component with few model assumptions. Furthermore eclipsing binaries provide a very good means to independently calibrate distance determination methods based on pulsating stars by comparison with the distance obtained from the binary star analysis. Until now only a few classical Cepheids were identified in eclipsing binaries (Pietrzy{\'n}ski et al. 2010, 2011) and one system with a pseudo RR Lyrae component was reported (Pietrzy{\'n}ski et al. 2012), there are also candidates for such systems that await confirmation.

The typical eclipsing binary star model consists of fixed size stars. To account for pulsations one usually: 1) modifies an existing modeling tool, 2) develops a new computer code, 3) removes pulsations from light and radial velocity curves and solves them with ordinary eclipsing binary star model. The first approach was used by Wilson \& van Hamme (2010) in case of the well-known Wilson-Devinney code (Wilson \& Devinney 1971; hereafter WD code), but only some phenomenological model was reported. The second approach was employed by MACHO to eclipsing binary Cepheids (Alcock et al. 2002, Lepischak et al. 2004), however the code was restricted only to light curve analysis and a rather simplistic treatment of stellar surfaces was used (e.g. no proximity and reflection effects were accounted for).  The third method seems to be the most common and is used in case of non-radial pulsators like $\delta$-Scuti stars (e.g. Southworth et al. 2011) or $\gamma$-Doradus stars (e.g. Maceroni et al. 2013).
The drawback of such an approach is that during eclipses pulsations can be removed only approximately from the light curves, which produces some systematic residuals in the solution. The way to partly overcome this difficulty is to employ the iterative light curve solution with the amplitude of the pulsations scaled according to the relative light contribution of a pulsating star during eclipses. This method was applied by Pietrzy{\'n}ski et al. (2010, 2011). To deal fully with changes in the eclipse geometry caused by pulsations, a novel approach is needed where both spectroscopic and photometric data are treated consistently.

We present here a new method of modeling eclipsing binaries with radially pulsating components. Instead of a new code development we decided to use a well-known and thoroughly tested computer model called JKTEBOP (Popper \& Etzel 1981, Southworth 2004, 2007) as a core of our method.  A Python based wrapper that we prepared can generate binary light curves with pulsations taken into account using original JKTEBOP code without any modifications. A similar methodology was proposed by Riazi \& Abedi (2006) in case of the WD code, but only for illustrative purposes.

The method was applied to the case of the eclipsing binary Cepheid OGLE-LMC-CEP-0227 (Pietrzy{\'n}ski et al. 2010, Soszy{\'n}ski et al. 2008) in the Large Magellanic Cloud. One of the main reasons to develop our approach was to directly determine the projection factor ($p$-factor) for the Cepheid from an eclipsing binary analysis. The $p$-factor is defined as the conversion factor between the observed pulsation radial velocities and the velocity of the pulsating star's photosphere. It plays a crucial role in Baade-Wesselink (Baade 1926, Wesselink 1946) type methods employed to pulsating stars like Cepheids. Its exact value and functional dependence on e.g. pulsation period is currently actively debated -- see Section~\ref{sub:factor} in this paper for references. In our opinion, the presented method allows for a robust determination of the $p$-factor for pulsating components of detached eclipsing binary systems.

Three groups were involved in the process of the preparation of this manuscript, namely the Araucaria project (data, software, analysis), the Carnegie Hubble Project (CHP; data) and the OGLE project (data). The research was based on observations obtained for ESO Programme 086.D-0103(A), 085.D-0398(A), 084.D-0640(A,B) and CNTAC time allocation 2010B-059.

\section{Data}
\label{sect:data}

Before starting the analysis we had to make sure that we have good enough data to obtain reliable results. When the discovery of the object was announced in 2008 by Soszy{\'n}ski et al. the eclipses were rather scarcely covered by the photometry and this analysis would not be possible. Since then as a result of a special observing program a strong emphasis was put on the measurement of the brightness change during eclipses.

\begin{table}
\caption{OGLE V-band photometry sample (the full version is available on-line). The errors are scaled to match the condition that the reduced $\chi^2$ is equal to 1.}
\centering
\begin{tabular}{c|c|c}
\hline
$HJD - 2450000$~d & V [mag] & error [mag] \\
\hline
3001.64990 & 14.984 & 0.008 \\ 
3026.74985 & 15.484 & 0.008 \\ 
3331.74155 & 15.023 & 0.008 \\ 
3341.74543 & 15.690 & 0.008 \\ 
3355.73941 & 15.347 & 0.008 \\ 
3359.66848 & 15.316 & 0.008 \\ 
3365.64874 & 15.419 & 0.008 \\ 
... & ... & ... \\ 
\hline
\label{tab:phot_v}
\end{tabular}
\end{table}

\begin{table}
\caption{OGLE I-band photometry sample (the full version is available on-line). The errors are scaled to match the condition that the reduced $\chi^2$ is equal to 1.} 
\centering
\begin{tabular}{c|c|c}
\hline
$HJD - 2450000$~d & I [mag] & error [mag] \\
\hline
2166.83748 & 14.353 & 0.007 \\ 
2172.88623 & 14.561 & 0.007 \\ 
2189.84343 & 14.365 & 0.007 \\ 
2212.79165 & 14.377 & 0.007 \\ 
2217.77657 & 14.492 & 0.007 \\ 
2223.79686 & 14.345 & 0.007 \\ 
2226.77167 & 14.303 & 0.007 \\ 
... & ... & ... \\ 
\hline
\label{tab:phot_i}
\end{tabular}
\end{table}

\begin{table}
\caption{Spitzer 3.6$\mu$m photometry sample (the full version is available on-line). The errors are scaled to match the condition that the reduced $\chi^2$ is equal to 1.} 
\centering
\begin{tabular}{c|c|c}
\hline
$HJD - 2450000$~d & 3.6$\mu$m [mag] & error [mag] \\
\hline
5813.54149 & 13.229 & 0.007 \\ 
5813.99999 & 13.264 & 0.007 \\ 
5814.57068 & 13.299 & 0.007 \\ 
5814.90875 & 13.262 & 0.007 \\ 
5815.55775 & 13.197 & 0.007 \\ 
5816.05859 & 13.225 & 0.007 \\ 
5816.58069 & 13.239 & 0.007 \\ 
... & ... & ... \\ 
\hline
\label{tab:phot_36}
\end{tabular}
\end{table}

\begin{figure}
\begin{center}
  \resizebox{\linewidth}{!}{\includegraphics{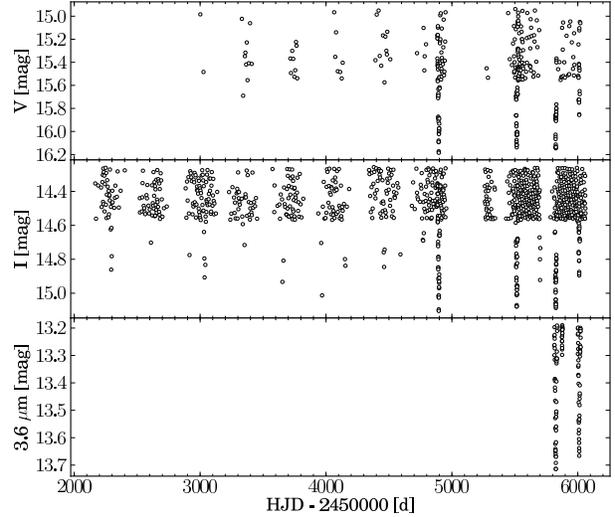}} \\
\caption{All photometric data collected for OGLE-LMC-CEP-0227. {\it Upper panel:} OGLE V-band data, {\it middle panel:} OGLE I-band data, {\it lower panel:} Spitzer $3.6 \mu m$ data. Note the difference in the eclipses coverage after the detection of the system at HJD about 2454900 days.
\label{fig:obs_all}}
\end{center}
\end{figure}

In total we acquired 1045 measurements in the I-band and 317 in the V-band collected with the Warsaw telescope by the OGLE project (Udalski 2003, Soszy{\'n}ski et al. 2012) and during the time granted to the Araucaria project by CNTAC organization. The auxiliary K-band data (only outside eclipses) were acquired by the Araucaria group using SOFI instrument attached to the NTT telescope at La Silla Observatory, which allowed us to use the V-K color variation to calculate the effective temperature as a function of the pulsation phase.  We have also acquired $3.6 \mu m$ and $4.5 \mu$ photometry from the Spitzer Space Telescope (114 points) -- the observations and data reduction provided by the CHP team. Because in the near-infrared the stellar limb darkening is low and the amplitude of the pulsations is small these observations put an important constraint on the geometry of the system. Fig.~\ref{fig:obs_all} presents all the photometric data used in our analysis. The {\it Spitzer} data were collected for two consecutive eclipses, i.e. for one primary and one secondary eclipse, and for one pulsation cycle outside the eclipses to obtain the unaffected pulsational light curve. In the analysis we used only $3.6 \mu m$ photometry because the out-of-eclipse observations in $4.5 \mu m$ band were too scarce and had the signal to noise ratio too low to obtain a correct representation of the pulsations (which could subsequently be used in the modeling) at this moment. We plan to complement the data in the future, however.
All the photometry used in this paper is provided in Tables \ref{tab:phot_v}-\ref{tab:phot_36} and in electronic form on:

\centerline{ http://araucaria.astrouw.edu.pl/p/cep227 }
\centerline{}

\begin{table*}
\caption{Radial Velocity Measurements of CEP-0227. HARPS spectra are marked with $^{\textstyle a}$ and UVES spectra are marked with $^{\textstyle b}$.} 
\begin{tabular}{@{}l|c|c|c|l|c|c|c|l|c|c|c@{}} \hline
HJD        & $RV_1$ & $RV_2$ & $RV_p$ & HJD        & $RV_1$ & $RV_2$ & $RV_p$ & HJD        & $RV_1$ & $RV_2$ & $RV_p$ \\
-2450000 d & (km/s) & (km/s) & (km/s) & -2450000 d & (km/s) & (km/s) & (km/s) & -2450000 d & (km/s) & (km/s) & (km/s) \\\hline
4810.76470$^{\textstyle a}$ & 288.44 & 223.94 &   3.40  &  5148.72671 & 292.06 & 219.71 &   3.50 &  5457.87258 & 293.69 & 220.22 &  10.72 \\
4811.58524$^{\textstyle a}$ & 288.94 & 223.57 &  15.08  &  5149.70121 & 291.62 & 219.80 &  17.26 &  5459.73087$^{\textstyle b}$ & 292.05 & 219.63 &  -2.22 \\
4854.60858 & 285.72 & 225.99 & -26.84  &  5149.81700$^{\textstyle b}$ & 292.10 & 220.04 &  18.69 &  5459.82242 & 292.77 & 220.81 &   0.19 \\
4854.76634 & 287.31 & 226.63 & -23.86  &  5150.66019$^{\textstyle b}$ & 291.49 & 219.98 & -26.65 &  5459.87734 & 292.35 & 220.74 &   0.84 \\
4855.62840 & 286.57 & 226.81 &  -7.73  &  5150.81070$^{\textstyle b}$ & 290.88 & 220.03 & -26.82 &  5459.88346$^{\textstyle b}$ & 291.38 & 219.94 &  -0.02 \\
4882.54381 & 270.06 & 243.88 &  -2.23  &  5151.62274$^{\textstyle b}$ & 291.74 & 220.89 & -12.04 &  5460.86428$^{\textstyle b}$ & 291.75 & 220.58 &  14.85 \\
5129.67036 & 292.48 & 220.78 &   3.26  &  5151.72685$^{\textstyle b}$& 291.48 & 220.84 &  -9.99 &  5461.71818$^{\textstyle b}$ & 291.33 & 220.31 &   1.93 \\
5129.68627 & 292.43 & 220.70 &   3.42  &  5152.62706$^{\textstyle b}$ & 291.15 & 220.70 &   5.38 &  5461.84896$^{\textstyle b}$ & 291.24 & 220.97 & -17.29 \\
5129.77828 & 292.57 & 220.69 &   4.90  &  5152.71553$^{\textstyle b}$ & 290.87 & 221.07 &   6.94 &  5462.74956$^{\textstyle b}$ & 291.19 & 220.87 & -17.54 \\
5129.79450 & 292.60 & 220.68 &   5.20  &  5155.60958 & 290.48 & 222.21 &  -7.87 &  5463.73892$^{\textstyle b}$ & 290.50 & 221.10 &   1.27 \\
5129.85284 & 292.72 & 220.89 &   6.39  &  5155.71877 & 290.45 & 222.26 &  -5.61 &  5463.87652$^{\textstyle b}$ & 290.72 & 221.30 &   3.56 \\
5130.67281 & 292.89 & 220.30 &  17.69  &  5155.71895 & 290.02 & 222.31 &  -6.03 &  5464.73657$^{\textstyle b}$ & 290.47 & 221.45 &  15.64 \\
5130.68789 & 292.62 & 220.38 &  17.69  &  5155.84086 & 290.82 & 222.69 &  -3.35 &  5464.87024$^{\textstyle b}$ & 290.30 & 221.44 &  17.73 \\
5131.67796 & 293.56 & 220.37 & -25.20  &  5167.78617$^{\textstyle b}$ & 285.18 & 227.84 &   5.46 &  5465.70513$^{\textstyle b}$ & 290.12 & 221.67 & -22.52 \\
5131.69389 & 293.32 & 220.18 & -25.51  &  5169.59889$^{\textstyle b}$ & 284.39 & 228.61 & -24.86 &  5465.81925$^{\textstyle b}$ & 289.79 & 221.63 & -26.61 \\
5131.78765 & 292.96 & 220.14 & -25.60  &  5174.84169$^{\textstyle b}$ & 280.30 & 231.55 &  -4.09 &  5466.70871$^{\textstyle b}$ & 289.60 & 222.23 & -14.11 \\
5131.80356 & 292.42 & 220.19 & -26.06  &  5185.53522$^{\textstyle a}$ & 273.67 & 237.70 & -18.20 &  5466.81183$^{\textstyle b}$ & 289.54 & 222.59 & -11.69 \\
5131.85265 & 293.03 & 220.25 & -25.19  &  5185.65259$^{\textstyle a}$ & 274.36 & 239.12 & -15.18 &  5467.75339$^{\textstyle b}$ & 289.34 & 222.65 &   4.92 \\
5131.86398 & 292.36 & 220.00 & -25.80  &  5185.79925$^{\textstyle a}$ & 273.48 & 238.98 & -12.38 &  5468.69682$^{\textstyle b}$ & 288.61 & 222.95 &  17.88 \\
5132.67972 & 292.99 & 220.00 & -10.91  &  5187.53794$^{\textstyle a}$ & 273.26 & 239.88 &  16.16 &  5468.85045$^{\textstyle b}$ & 290.13 & 222.85 &  20.08 \\
5132.69356 & 292.98 & 220.06 & -10.62  &  5187.68699$^{\textstyle a}$ & 271.54 & 239.89 &  16.97 &  5469.68392$^{\textstyle a}$ & 289.74 & 224.34 & -25.09 \\
5132.69358 & 293.54 & 220.02 & -10.05  &  5187.80119$^{\textstyle a}$ & 272.91 & 240.05 &  19.16 &  5477.79364$^{\textstyle a}$ & 283.78 & 227.69 & -19.80 \\
5132.76078 & 293.15 & 220.21 &  -9.00  &  5251.52915 & 238.30 & 273.28 &   8.94 &  5477.87033$^{\textstyle a}$ & 284.65 & 227.85 & -18.25 \\
5132.77322 & 293.25 & 220.18 &  -8.64  &  5251.52920 & 238.60 & 273.91 &   9.24 &  5478.87548$^{\textstyle a}$ & 283.67 & 228.16 &   0.45 \\
5132.85291 & 293.04 & 220.14 &  -7.12  &  5251.59065 & 238.78 & 274.38 &  10.58 &  5479.78887$^{\textstyle a}$ & 283.59 & 228.98 &  14.66 \\
5132.86601 & 293.05 & 220.40 &  -6.84  &  5251.64129 & 238.57 & 274.67 &  11.27 &  5479.87635$^{\textstyle a}$ & 284.03 & 228.66 &  15.73 \\
5132.87916 & 292.97 & 220.19 &  -6.64  &  5251.64131 & 238.84 & 274.60 &  11.54 &  5502.84368$^{\textstyle a}$ & 269.47 & 243.35 &  18.43 \\
5141.77686$^{\textstyle b}$ & 293.27 & 219.22 &  14.23  &  5251.69466 & 238.54 & 274.81 &  12.14 &  5559.58184$^{\textstyle b}$ & 239.06 & 272.86 &  14.74 \\
5144.62452$^{\textstyle a}$ & 292.97 & 219.09 &  -1.41  &  5251.69469 & 238.96 & 274.91 &  12.56 &  5560.80361$^{\textstyle b}$ & 238.73 & 274.02 & -26.25 \\
5144.68958$^{\textstyle a}$ & 293.58 & 219.16 &   0.46  &  5251.74836 & 238.41 & 275.08 &  12.86 &  5561.81170$^{\textstyle b}$ & 237.77 & 273.55 & -10.43 \\
5144.76237$^{\textstyle a}$ & 293.90 & 219.11 &   2.15  &  5272.50251 & 232.83 & 280.22 & -23.32 &  5582.55697$^{\textstyle b}$ & 232.90 & 280.11 &  18.03 \\
5145.64671$^{\textstyle a}$ & 292.38 & 219.63 &  13.99  &  5272.55269 & 232.64 & 280.70 & -22.32 &  5583.56084$^{\textstyle b}$ & 231.99 & 280.22 & -26.60 \\
5145.72154$^{\textstyle a}$ & 294.21 & 219.98 &  16.28  &  5272.63451 & 231.82 & 280.75 & -21.27 &  5584.64151$^{\textstyle b}$ & 232.31 & 280.73 &  -8.55 \\
5146.61150$^{\textstyle a}$ & 291.77 & 220.04 &  -7.11  &  5272.63454 & 232.47 & 281.15 & -20.61 &  5588.59591$^{\textstyle b}$ & 231.20 & 281.25 &  -5.65 \\
5146.67262$^{\textstyle a}$ & 292.15 & 220.02 & -15.48  &  5272.68201 & 233.08 & 281.10 & -19.19 &  5590.53541 & 231.57 & 282.04 &  22.62 \\
5146.73621$^{\textstyle a}$ & 292.75 & 219.22 & -21.43  &  5272.68214 & 233.53 & 281.42 & -18.74 &  5590.58281 & 230.76 & 281.71 &  22.36 \\
5146.79951$^{\textstyle a}$ & 293.61 & 219.38 & -24.13  &  5431.78359$^{\textstyle a}$ & 290.07 & 223.10 & -25.14 &  5590.63837 & 230.56 & 282.11 &  21.27 \\
5147.55995$^{\textstyle b}$ & 294.12 & 220.78 & -16.46  &  5431.82587$^{\textstyle a}$ & 288.39 & 223.70 & -26.60 &  5590.63846 & 230.56 & 281.51 &  21.26 \\
5147.69997$^{\textstyle b}$ & 293.16 & 219.99 & -14.64  &  5431.88740$^{\textstyle a}$ & 290.85 & 222.81 & -23.66 &  5590.68403 & 231.89 & 282.54 &  20.08 \\
5147.83139$^{\textstyle b}$ & 292.56 & 219.89 & -12.01  &  5457.75868 & 291.65 & 220.75 &  19.98 &  5590.72419 & 232.35 & 283.02 &  16.82 \\
5148.63334$^{\textstyle b}$ & 292.19 & 219.97 &   2.27  &  5457.81626 & 292.94 & 220.54 &  16.88 &  5598.66755$^{\textstyle b}$ & 230.19 & 282.81 & -25.09 \\\hline
\label{tab:vel}
\end{tabular}
\end{table*}

The photometry alone, however, is not sufficient to obtain the absolute values of some important parameters like mass or scale of the system. Using the MIKE spectrograph at the 6.5-m Magellan Clay telescope at Las Campanas Observatory in Chile, the HARPS spectrograph attached to the 3.6-m telescope at La Silla Observatory and the UVES spectrograph on VLT at Paranal Observatory we obtained 123 high-resolution spectra at 116 epochs (49 MIKE + 27 HARPS + 40 UVES) -- 76 more than those reported in Pietrzy{\'n}ski et al. (2010). All the observations were performed by the Araucaria project. Using these data we also confirmed the OGLE-LMC-CEP-0227 (hereafter CEP-0227) to be a classical fundamental-mode Cepheid pulsator in a well detached, double-lined, eclipsing system. The object turned out to have near-perfect properties for deriving the masses of its two components with a very high accuracy. 

Radial velocities (RVs) of both components were measured using RaveSpan application (Pilecki et al. 2012). We have used the Broadening Function formalism (Rucinski 1992, 1999) with templates matching the stars in the temperature-gravity plane. The templates were theoretical spectra taken from the library of Coelho et al. (2005). For deriving radial velocities we have analyzed the spectra in the range of 4125 to 6800 $\mathring{A}$. The typical formal errors of the derived velocities are $\sim300$ m/s. In case of a Cepheid component the orbital motion had to be extracted from the original radial velocities by the subtraction of the pulsational RVs. We assumed: a mutual Keplerian motion of both components with a constant orbital period, that the stars were point-like sources (i.e. no proximity effects, like star's oblateness, were incorporated at this stage of analysis) and that the pulsation radial velocity curve could be represented by the Fourier series. We fitted simultaneously orbital period $P$, eccentricity $e$, periastron longitude $\omega$, both velocity semi-amplitudes $K_1$ and $K_2$, both stars systemic velocities $\gamma_1$ and $\gamma_2$, and a number of Fourier series coefficients (depending on the series order).  In Table~\ref{tab:vel} the orbital radial velocities of both components, $RV_1$ and $RV_2$, together with the pulsation radial velocities of the Cepheid $RV_p$ are presented. The original Cepheid radial velocities are simply $RV=RV_1+RV_p$. The analysis confirmed the presence of the K-term effect (Nardetto et al. 2008 and references therein), that affects Cepheid-type stars: the Cepheid systemic velocity is blue-shifted in respect to the companion systemic velocity by 0.59 km/s.

\section{Method}
\label{sect:method}
\subsection{Light curve synthesis}
\label{sub:lit}
As far as we know there is no generally available software that allows to model binary eclipsing stars with pulsating components in a fully consistent physical way. So first, we developed a scheme which was later on followed by the software application that allows the standard modeling tools, like WD or JKTEBOP code, to model this kind of systems.

The trick is to generate multiple eclipsing light curves for different stages of a pulsating component while the parameters of the pulsating star remain fixed for every single light curve generated.  This way we obtain a two-dimensional light curve that depends on both pulsational and orbital phase.
Later on for every observation point both orbital and pulsation phases are calculated and used to obtain a corresponding brightness from the 2D grid. A bilinear interpolation is used to calculate the brightness between the grid points to improve the efficiency and accuracy of the method.

\begin{figure}
\begin{center}
  \resizebox{0.92\linewidth}{!}{\includegraphics{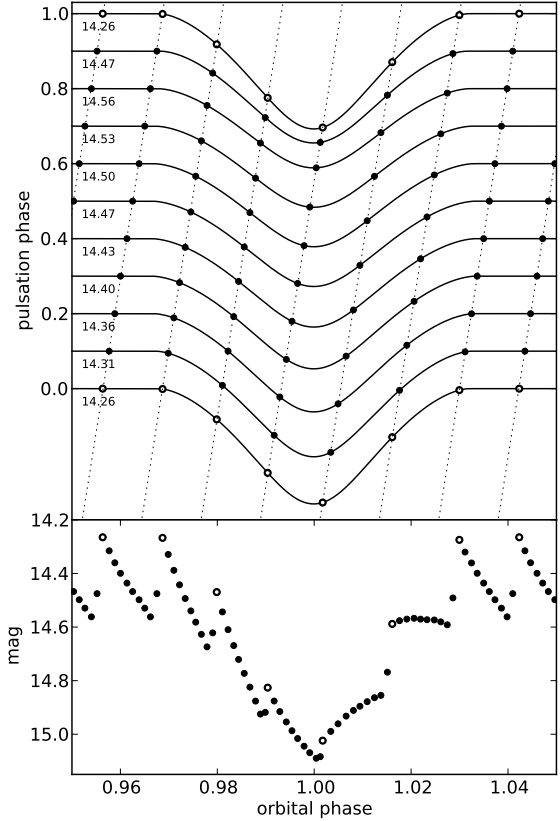}} \\
\caption{Generation of a 1D light curve (lower panel) from the 2D grid of light curves. {\it Upper panel:} this plot shows a small subset (just 10) of fixed-radius eclipsing models for different phases of a pulsating component centered on a primary minimum. The size and shape of the eclipses changes as we move through the Y-axis. Small numbers on the left side are the brightness of the system at the maximum. The pulsation maxima are marked with open circles. Diagonal lines mark the propagation of the pulsation phase as we move through the orbital phase -- the pulsation period is many times shorter than the orbital one. {\it Lower panel:} the resulting light curve when a pulsating component is obscured by a companion.
\label{fig:pri_ecl}}
\end{center}
\end{figure}

\begin{figure}
\begin{center}
  \resizebox{0.92\linewidth}{!}{\includegraphics{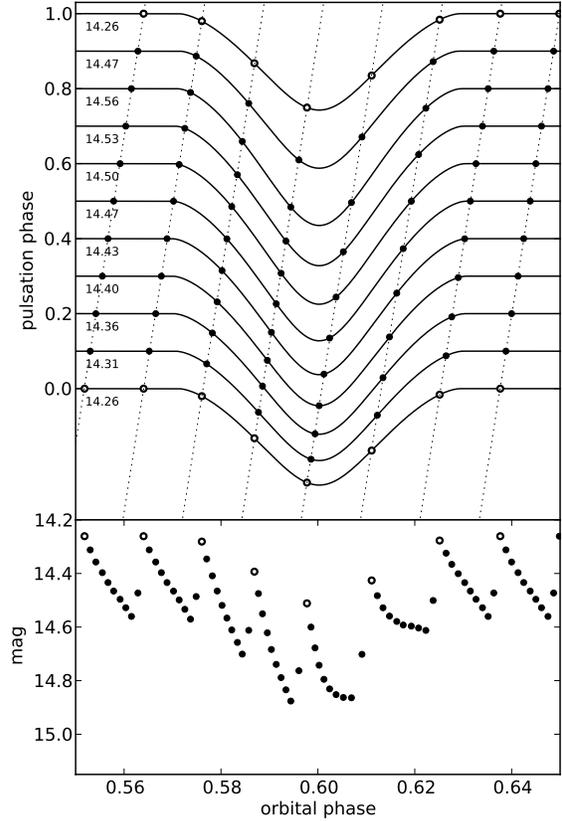}} \\
\caption{Same plot as in Fig.~\ref{fig:pri_ecl} but for the secondary eclipse. {\it Lower panel:} the resulting light curve when a pulsating component passes in front of its companion.
\label{fig:sec_ecl}}
\end{center}
\end{figure}

\begin{enumerate}

\item Generation of the 2-dimensional light curve.

For N uniformly spaced phases ($N=100$ in our approach) of the pulsation cycle we calculate the full eclipsing model using the JKTEBOP code. The generated light curve consists of $M=2000$ points uniformly covering the orbital cycle.
This number comes from a compromise between the accuracy of modeling the minima shape and the numerical efficiency of the code. In calculating the N models we take into account the following pulsation phase {\it dependent} parameters: the fractional radius of the primary $r_1$, the surface brightness ratio of the components $j_{21}$ and the brightness of the system in a given band (the light scale factor expressed in magnitudes). And the following pulsation phase {\it independent} parameters are kept fixed: the fractional radius of the secondary $r_2$, the eccentricity $e$, the periastron longitude $\omega$, the orbital inclination $i$, the mass ratio $q=m_2/m_1$ and the reference time of the primary minimum $T_0$. The fractional radii are the physical radii, $R_1$ and $R_2$, divided by the semi-marjor axis $a$. The pulsation period is kept constant and the pulsation phase is calculated according to the following ephemeris:
\begin{equation}
\label{pul:efe}
T_{max} ($HJD$) = 2454896.285 + 3.797086\times $E$,
\end{equation}
where $T_{max}$ refers to the moment of the Cepheid's maximum optical brightness. The reflection and proximity effects are taken into account internally by the JKTEBOP code, but in the case of CEP-0227 (being on the order of 0.001 mag) they are of minor importance.
As a result we obtain a two-dimensional grid of magnitudes $m = m(\phi_{\rm orb}, \phi_{\rm puls})$ for each photometric band ($V$, $I_C$ and Spitzer $3.6 \mu$m) we use.

The radius of the pulsating component changes during the pulsation cycle. To account for this effect we use the Cepheid disentangled pulsational radial velocities and the projected semi-major axis of the system $a\sin{i}$ -- see details in Section~\ref{sub:radius}.

\begin{figure}
\begin{center}
  \resizebox{\linewidth}{!}{\includegraphics{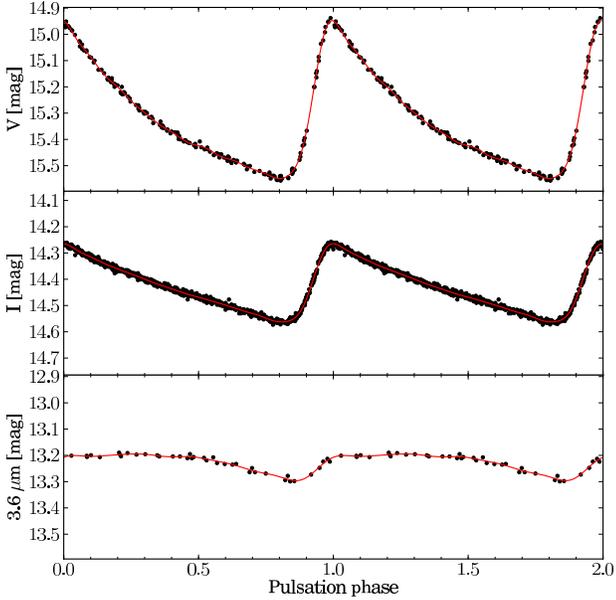}} \\
\caption{Out-of-eclipse light curves of the CEP-0227 folded with the ephemeris given in equation~(\ref{pul:efe}). The overplotted 9th (V and I) and 6th (3.6$\mu$m) order Fourier fits (solid lines) are used to calculate the light scale factor expressed in magnitudes for a given pulsation phase which is an input parameter of JKTEBOP. 
\label{fig:puls}}
\end{center}
\end{figure}

The light scale factor is calculated from the out-of-eclipse light curve for a given pulsation phase. To parameterize the pulsations we use 9th order Fourier series for V and I bands and a 6th order one for the Spitzer data. Next, this fit is used to set the light scale factor for all N pulsation phases in every band. These out-of-eclipse light curves and Fourier fits are presented in Fig.~\ref{fig:puls}. 

The method to calculate the actual surface brightness ratio of the components is presented in Section~\ref{sub:surf}. The limb darkening coefficients were treated with special attention and we worked out the methodology to treat them in a consistent way within the model, for details see Section~\ref{sub:limbmet} of this paper. Reflection coefficients were calculated from the model geometry. The gravitational brightening was set to 0.32 -- a value typical for a convective envelope of a late type star.

\item Creation of 1D light curve from the 2D one:

Once all purely eclipsing models (without pulsations) are calculated for a set of different pulsation phases (from 0.0 to 1.0) we are ready to calculate a 1D light curve that exhibits pulsations. For this we need the specific epochs of observation, because the pulsational and eclipsing variabilities are independent. Thus for each measurement we calculate the both phases and take from the 2D grid the interpolated value that best represents the actual brightness of the system in a given photometric band. This allows us to obtain a direct eclipsing binary light curve model with a pulsating component like the one in Fig.~\ref{fig:pri_ecl} and \ref{fig:sec_ecl}. Note that those figures were simplified to present the idea with more clarity. In reality we do not use brightness values from where the diagonal lines cross the light curve, we use the brightness values for the exact (calculated) pulsation phases. As for the pulsation phase, there is also a small correction applied at this moment during the selection of the best model from the grid, due to the light time travel effect (see Section~\ref{sub:ltte} for details).

\end{enumerate}

In the way described above we manage to obtain a light curve model for a given set of orbital and stellar parameters. In order to find the set of parameters giving the best fit to the observations we decided to employ Markov chain Monte Carlo (MCMC) approach (Press et al. 2007). To fully explore the parameter space we allow for the change of the following parameters: the fractional radius of the pulsating component at phase 0.0 (pulsational), the fractional radius of the second component, the orbital inclination, the orbital period, the reference time of the primary minimum, the eccentricity related parameters ($e\cos\omega$, $e\sin\omega$), the component surface brightness ratios in all the used bands at phase 0.0 (pulsational), the $p$-factor and the third light $l_3$. 

\subsection{Markov chain Monte Carlo}
\label{sub:mcmc}

We decided to use the Monte Carlo method as it allowed us to realistically estimate errors of the parameters. Specifically we have used Metropolis-Hastings algorithm (Hastings 1970) -- one of the MCMC random-walk methods, which has an advantage over non random-walk MC sampling being in general independent of the starting point (unless you start it close to the other deep local minimum) and sampling the $\chi^2$ plane with greater density where the $\chi^2$ values are lower. As the acceptance function we use the normal distribution function. The method was also modified by the incorporation of simulated annealing (Press et al. 2007) -- the probability of jumping away from the $\chi^2$ minimum decreases as the number of calculated models increases. We would like to emphasize here that we fit all the light curves simultaneously i.e. geometry related parameters like radii, orbital inclination, $p$-factor, etc. are common to all bands. The observations are weighted by their observational errors and their modal values are 0.009, 0.007 and 0.008 mag in $V$, $I_C$ and 3.6$\mu$m, respectively. At the initial stage all errors were scaled to match the condition that for every single light curve the reduced $\chi^2$ should be equal to unity.

To obtain the well-sampled $\chi^2$ plane for 12 fitted parameters we need about 50 000 models to be calculated. While 10 000 gives a good estimate of the best solution, at least 5 times more models is needed to reliably estimate the errors.

\subsection{Radius change}
\label{sub:radius}
\begin{figure}
\begin{center}
  \resizebox{\linewidth}{!}{\includegraphics{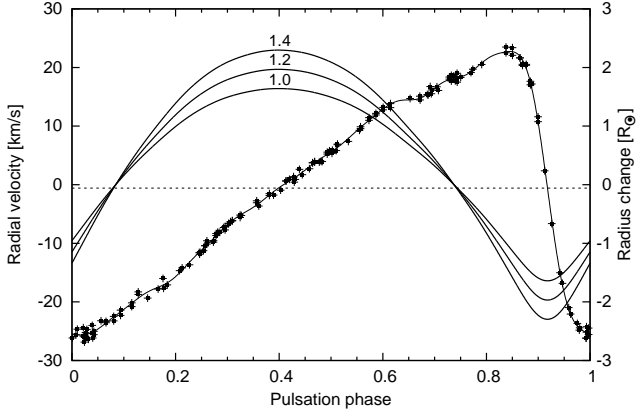}} \\
\caption{Pulsational radial velocities of the Cepheid OGLE-LMC-CEP-0227 (crosses) with the over-plotted 12th order Fourier series fit. The dashed straight line corresponds to the Cepheid's systemic velocity of $-0.59$ km/s in respect to the barycenter of the system. Three continuous lines correspond to the Cepheid's radius changes, in respect to the mean radius, for the projection factor values of 1.0, 1.2 and 1.4. 
\label{fig:pulsrad}}
\end{center}
\end{figure}

The radius absolute change of the pulsating component can be found directly from integrating the pulsation radial velocity curve:

\begin{equation} \Delta R_1 (t,p) = B \!\int \!\!p\,(v_r(t) - v_{s})\, {\rm d}t  = p D(t),
   \label{eqn:radchange}
   \end{equation}
where $p$ is the projection factor, $v_r$ is the radial velocity, $B$ is a conversion unit factor and $v_s$ is the Cepheid systemic velocity with respect to the system barycenter. If we choose units to be solar radii for length, km s$^{-1}$ for velocity and days for time we get $B=0.12422$. The systemic velocity $v_s$ is selected to give a zero net effect of the radius change after a pulsation cycle, i.e. we require the star to have always the same radius at a given phase. Fig.~\ref{fig:pulsrad} shows the pulsational radial velocity curve used in the analysis and the resulting radius changes.

In general the projection factor can be phase dependent, but as shown by Nardetto et al. (2004) the impact of this dependence is weak ($0.2\%$ on the distance determination) and in our analysis we keep it constant for a given model (it is not fixed in regard to the MCMC analysis though). It is convenient to separate the time independent $p$-factor and the parameter independent D(t) term, as the latter can be calculated once for the whole analysis.
Let us denote the Cepheid fractional radius and its absolute radius correction at pulsation phase 0.0 by $r_{1,0}$ and $\Delta R_{1,0}$, then the fractional radius of the pulsating component at any time can be found from the relation:

\begin{equation}   r_1(t,p) = r_{1,0} + \frac{\Delta R_1(t,p) - \Delta R_{1,0}}{a},
   \label{eqn:radius}
   \end{equation}
where $a$ is the semi-major axis of the system obtained from the orbital solution -- see Section~\ref{sub:orbsol}.

\subsection{Surface brightness ratio}
\label{sub:surf}
The dimensionless surface brightness ratio of the components that influences the depth of the eclipses is one of the fitted parameters within the JTKEBOP code. This quantity changes during the pulsation cycle because the effective temperature of the Cepheid is not constant and, what follows, neither is the mean flux emitted from the surface area of the star. To calculate the surface brightness ratio in a given moment and band we use an adequate out-of-eclipse pulsation light curve and the fractional radius changes given by equation~\ref{eqn:radius}. Let us denote the Cepheid flux, its surface brightness and the total apparent brightness of the system measured in a given band at pulsation phase 0.0 by $F_{1,0}$, $j_{1,0}$ and $m_0$, respectively. Their current values during pulsations are $F_{1}(t)$, $j_{1}(t)$ and $m(t)$. The radius of the secondary component $r_{2}$ and its surface brightness $j_{2}$ are constant during the pulsation cycle. Let us now define the surface brightness ratio at pulsation phase 0.0 and its current value by:
\begin{eqnarray} j_{21,0}&=&\frac{j_{2}}{j_{1,0}} \nonumber \\
 j_{21}(t)&=&\frac{j_{2}}{j_{1}(t)}
\label{eqn:surfdef}
\end{eqnarray}
In general some amount of the third light in the system, being an optical blend or an additional physical stellar companion, may be present. Although it does not affect physically the surface brightness ratio of the components, it affects the way we derive this quantity from the out-of-eclipse magnitudes. Let's assume that the third light flux is constant in a given band $F_{3}=const$. Then we can define the third light $l_3$, which is one of the input parameters to the JKTEBOP code, by:
 \begin{eqnarray} l_{3,0}&=&\frac{F_{3}}{F_{1,0}+F_2+F_3} \nonumber \\
 l_{3}(t)&=&\frac{F_{3}}{F_{1}(t)+F_2+F_3},
\label{eqn:thirdlight}
\end{eqnarray}   
 where $F_2$ is the flux from the companion and $l_{3,0}$ is the third light at phase 0.0. Note, that although we assume constant $F_3$, the third light $l_3$ changes during the pulsation cycle because its contribution to the total light changes. By the Pogson equation we can link the instantaneous and reference apparent brightness $m(t)$ and $m_0$:
\begin{equation} m(t) - m_0 = -2.5 \log \frac{F_1(t)+F_2+F_3}{F_{1,0}+F_2+F_3}
\label{eqn:pogson}
\end{equation}

From equation~(\ref{eqn:thirdlight}) we derive $F_3$:
\begin{equation} F_3 = \frac{l_{3,0}(F_{1,0}+F_2)}{1-l_{3,0}}
\label{eqn:flux3}
\end{equation}
The fluxes from both components $F_1$ and $F_2$ are proportional to the product of their projected surface area and surface brightness:
 \begin{eqnarray} F_1(t)&\sim& r_1^2(t)\,j_1(t) \nonumber \\
F_{1,0}&\sim& r_{1,0}^2\,j_{1,0} \nonumber \\
F_2&\sim& r_2^2\,j_2 
\label{eqn:fluxes}
\end{eqnarray} 
where $r_1$ dependence on $p$ is omitted as for any given model the $p$-factor is constant across the pulsation cycle. Inserting equations~(\ref{eqn:flux3}) and (\ref{eqn:fluxes}) into equation~(\ref{eqn:pogson}), and after some algebraic manipulations with the help of equation~(\ref{eqn:surfdef}) we obtain a pulsation phase dependent relation for the surface brightness ratio:
\begin{equation} j_{21}(t) = \frac{r_1^2(t)\,j_{21,0}}{(r_{1,0}^2+r_2^2\,j_{21,0})\left(\frac{\textstyle A(t)}{\textstyle 1-l_{3,0}} - l_{3,0}\right) - r_2^2\,j_{21,0}}
\label{eqn:surfbrit}
\end{equation}
where $A(t)=10^{0.4(m_0-m(t))}$. Solving equations~(\ref{eqn:thirdlight}) and~(\ref{eqn:pogson}) for $l_3(t)$ we obtain an expression for the phase dependency of the third light parameter:
\begin{equation} l_3(t) = \frac{l_{3,0}}{{\textstyle A(t)}}
\label{eqn:3t}
\end{equation}

\subsection{Limb darkening methodology}
\label{sub:limbmet}
The limb darkening (LD) of a star surface affects determination of stellar radii in case of the eclipsing binary light curve analysis. Instead of fitting the LD coefficients in all three photometric bands independently we decided to link them by atmospheric parameters i.e. the effective temperature $T_{eff}$, the gravity $\log g$ and metallicity $[$Fe/H$]$ utilizing some theoretical LD predictions. We have tested two sets of stellar limb darkening tables published by Van Hamme (1993) and Claret \& Bloemen (2011) and two limb darkening laws, namely a linear and a logarithmic one (Klinglesmith \& Sobieski 1970). As the Van Hamme tables lack data for the Spitzer $3.6\mu$m band, equivalent Johnson $L$ band coefficients were used instead.

Because atmospheric parameters of Cepheids change during pulsation cycle we expected that also limb darkening coefficients would change over time. To account for this effect we need to know how both parameters vary with pulsation phase.  
The instantaneous surface gravity of the Cepheid is calculated from:
$$ \log g (t) = 4.438 + \log{m_1} - 2 \log(r_1(t)\,a), $$
where $m_1$ is the mass of the Cepheid, $a$ is the orbital semi-major axis and the instantaneous star radius is calculated from equation~\ref{eqn:radius}. The masses of both components are adopted from the solution obtained with the WD code.   

\begin{figure}
\begin{center}
  \resizebox{\linewidth}{!}{\includegraphics{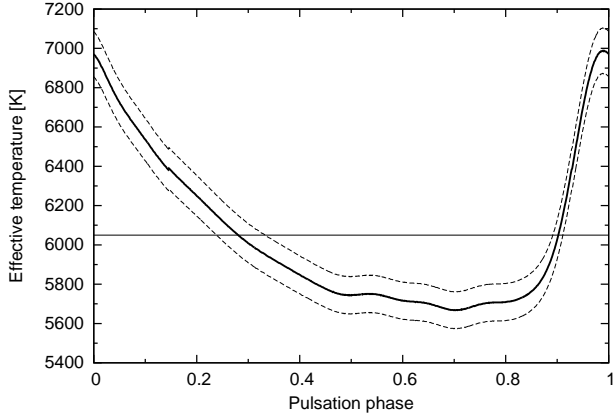}} \\
\caption{ Dependence of the Cepheid effective temperature over the pulsation phase (thick solid curve) with 1-$\sigma$ uncertainties (dashed lines). The temperature variation may be used to calculate limb darkening coefficients for a given phase.  The horizontal line represents the mean effective temperature ($6050$ K) of the star.
\label{fig:temp}}
\end{center}
\end{figure}

The effective temperature of the Cepheid can be inferred from the color indices like ($V\!-\!I$) or ($V\!-\!K$). In practice temperature calibrations based on ($V\!-\!K$) are much more reliable. In order to obtain intrinsic ($V\!-\!K$) colors of the Cepheid during the pulsation cycle we have to remove the light contribution of the accompanying red giant in $V$ and $K$ bands, and the same is needed to estimate the amount of interstellar reddening in the direction of CEP-0227. Once intrinsic ($V\!-\!K$) index is obtained the effective temperature is estimated using various calibrations. Details of this procedure are described in a separate paper (Gieren et al. in preparation). In Fig.~\ref{fig:temp} we present how the temperature of the Cepheid changes over one pulsation period. The metallicity $[$Fe/H$]=-0.5$ was assumed for both components. It is slightly larger than $[$Fe/H$]\sim-0.65$ derived by Marconi et al. (2013) but resulting change in the LD coefficients is insignificant. 

In case of the second component the effective temperature and gravity are constant, thus limb darkening coefficients do not need any special treatment. For the secondary component we set the constant effective temperature $T_{eff,2} = 5120$ K and gravity $\log g_2 = 1.71$. 

During our analysis it appeared that the limb darkening coefficients calculated for constant average effective temperature gave better results than those for the variable one. Therefore we tested this option thoroughly and eventually this was the main method that we have used. Note that it does not mean that the Cepheid temperature is constant nor it means that the LD coefficients are such. It only means that the dependence on the $T_{eff}$ may be different than the one assumed here. Indeed Marengo et al. (2003) based on the theoretical considerations found some significant variability of limb darkening between the pulsation phase $\phi=0.6$ and 0.7 coinciding with a shockwave passage through the photosphere. However for most part of the pulsation period LD coefficients were found to change only a little. 

\subsection{Light time travel effect}
\label{sub:ltte}
A pulsating Cepheid star is a kind of a cosmic clock. When a star orbits another star we observe that this clock is accelerating when a Cepheid is approaching us and decelerating when a Cepheid is drifting away.  This well known light time travel effect is no doubt present in the binary CEP-0227. The only question is if the photometry we collected is of the precision good enough to make this effect detectable.
The instantaneous distance $\rho$ of the Cepheid star from the system barycenter is given by equation:
\begin{equation}  \rho = \frac{a_1\,(1-e^2)}{1+e\cos\nu},
\label{eqn:kepler1}
\end{equation}
where $a_1$ is the semi-major axis of the Cepheid orbit, $e$ is the eccentricity and $\nu$ is the true anomaly. The projection of $\rho$ onto the line of sight passing through the system barycenter equals:
\begin{equation}  d = \rho\,\cos(\omega+\nu-\pi/2)\sin i
\label{eqn:kepler2}
\end{equation} 
where $\omega$ is the periastron longitude and $i$ is the orbital inclination. The value of $d$ tells us how much the Cepheid is closer or farther away from us in respect to the system barycenter. The time which light needs to pass this distance is a retardation of the pulsating signal.  In other words the observed pulsation phase $\phi_p^{\prime}$ is different from the pulsation phase $\phi_p$ computed for the constant mean pulsation period $P_p$ and they are related as follows:
\begin{equation}  
\label{eqn:light}  \phi_p^{\prime} =\phi_p - \frac{d}{cP_p},
\end{equation}    
where $c$ is the velocity of light. 

Although the retardation of the pulsation signal in the primary minimum is just $\sim0.0014$ of the pulsation period, during the most steep part of the pulsation light curve (between phases 0.85 and 1.0), it translates into 0.003 mag shift in $I$ band and 0.005 mag shift in $V$ band. Such shifts are comparable to the precision of OGLE-IV photometry and, being a systematic effect, can affect our solution. Indeed after implementation of the effect in our code we detected it on 3.5-$\sigma$ significance level. The implementation is made by applying a correction to the calculated pulsation phase while the best model is being taken from the 2D light curve grid.

\section{Results}
\label{sect:results}

As the initial parameters for our analysis we used the results from our previous study of this system (Pietrzy{\'n}ski et al. 2010). Because we have gathered a lot of new data and applied more sophisticated and direct approach, we expect the results to be more reliable and accurate. Some other effects neglected before were also taken into account this time.

First we have obtained a new orbital solution, which was then used as a base for the following analysis of the photometry using the method described above.

\subsection{Orbital solution}
\label{sub:orbsol}

\begin{table}
\caption{Orbital solution for CEP-0227. In RaveSpan stars are treated as point like sources, $T_0$ is $HJD - 2450000$~d, and $a \sin i$ is calculated with the rest frame orbital period $P=309.404$~d.} 
\begin{tabular}{@{}l|@{}c@{}|ccc@{}} \hline
 Parameter & RaveSpan & \multicolumn{2}{c}{WD} \\
 \multicolumn{2}{c}{} & Solution 1 & Solution 2 \\\hline 
$\gamma$ (km/s)       & 256.61 $\pm$ 0.04 &256.48  $\pm$ 0.11 &  256.46 $\pm$ 0.09 \\ 
$T_0$ (d)             &                   &4818.94 $\pm$ 0.28 & 4820.88 $\pm$ 0.42 \\
$a \sin i$ ($R_\odot$)& 384.24 $\pm$ 0.67 &389.26 $\pm$ 0.44 & 388.89 $\pm$ 0.77\\ 
$q=M_2/M_1$           & 0.993 $\pm$ 0.002 &0.993 $\pm$ 0.003& 0.994 $\pm$ 0.003\\ 
$e$                   & 0.163 $\pm$ 0.002 &0.166 (fixed) & 0.161 $\pm$ 0.003\\
$\omega$ (deg)        & 343.0 $\pm$ 1.4 &342.0 (fixed) & 344.5 $\pm$ 1.8\\
$K_1$ (km/s)          & 31.72 $\pm$ 0.06 & 32.14 $\pm$ 0.05 & 32.11 $\pm$ 0.07\\ 
$K_2$ (km/s)          & 31.94 $\pm$ 0.06 & 32.38 $\pm$ 0.05 & 32.31 $\pm$ 0.06\\ 
rms$_1$ (km/s)        & 0.54 & 0.56 & 0.55\\
rms$_2$ (km/s)       & 0.48 & 0.48 & 0.44\\\hline
 \label{tab:spec}
\end{tabular}
\end{table}

\begin{figure}
\begin{center}
  \resizebox{\linewidth}{!}{\includegraphics{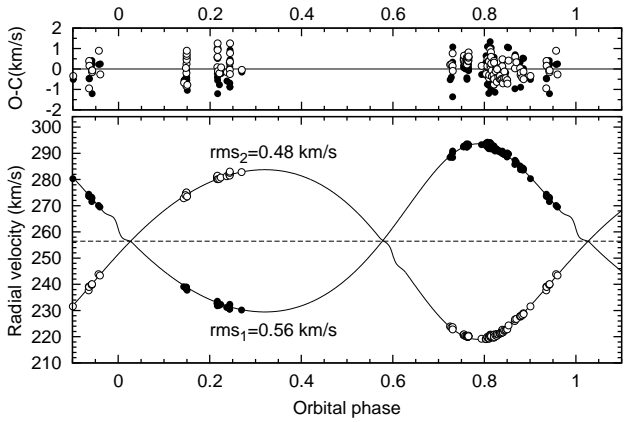}} \\
\caption{Orbital solution for CEP-0227 (solid lines). Measured radial velocities of the Cepheid with the pulsations removed (filled circles) and of its constant companion (open circles) are presented. Model residua are shown in the upper panel. 
\label{fig:specsol}}
\end{center}
\end{figure}

We analyzed disentangled orbital radial velocities of both components with Wilson-Devinney code (Wilson \& Devinney 1971, Wilson 1979, Van Hamme \& Wilson 2007) to derive the projected semi-major axis of the system $a\sin{i}$ and the mass ratio $q$. The reason behind using the WD code is to account for non-Keplerian corrections originating from stars oblateness. These corrections are relatively small ($\sim 0.4$ km/s) but result in the $a\sin{i}$ different by $3\sigma$ from the purely Keplerian solution (RaveSpan). The adjusted parameters were: the semi-major axis $a$, the systemic velocity $\gamma$, the mass ratio $q$ and the phase shift $\Delta\phi$. The remaining spectroscopic parameters were kept constant during fitting and their values were adopted from the photometric solution -- see Section~\ref{sub:photo}.  The velocity semi-amplitudes were calculated according to:
 \begin{eqnarray} K_2 [{\rm km/s}]&=& 50.579\frac{a\sin i[R_\odot]}{P[d]\,(1+q)\sqrt{1-e^2}} \\
K_1 [{\rm km/s}]&=&q\,K_2  
\label{eqn:semiamp}
\end{eqnarray} 

The results of the preliminary fitting with RaveSpan and the final fitting with the WD code are summarized in Table~\ref{tab:spec}. We also perform another run of the WD code adjusting the eccentricity $e$ and the periastron longitude $\omega$ to check the consistency of the photometric and spectroscopic solutions (Solution 2). The resulting $q$ and $a$ are essentially the same as for the one when $e$ and $\omega$ were kept constant. For later analysis we adopted the results from Solution~1 with error on semimajor axis from Solution~2.

\subsection{Limb darkening}

From the preliminary analysis we have learned that using the Van Hamme tables and the logarithmic law results in smaller residuals than for any other combination of tables and laws used, so they were selected for later analysis.
Also, as described in Section~\ref{sub:limbmet} for the primary pulsating component two scenarios have been considered: 1) limb darkening coefficients dependent on the $T_{eff}$ and $\log g$ (which change over pulsation phase), 2) limb darkening coefficients calculated for a constant $T_{eff}$ and variable $\log g$. The latter approach gave significantly better results in terms of $\chi^2$ values so we decided to use it to obtain a final solution.

Having set this we then varied $T_{eff}$ in order to find out how $\chi^2$ of the best solution depends on limb darkening. Surprisingly we had to lower the Cepheid temperature  (what corresponded to larger LD coefficients) to as low as $T_{LD,1} = 3700$ K. The improvement in $\chi^2$ was considerable and significant to 6-$\sigma$ level -- see Fig.~\ref{fig:t1chi2p}.
The minimum lies well within $1\sigma$ from the lower boundary for the tables used, which is 3500 K, but it appears that further decreasing of the temperature (i.e. increasing the LD coefficients) would not improve the fit. The final scaling factor for the temperature is $a_1 = 3700 K / 6050 K \approx 0.61$.

We have also tried to find a better solution varying LD coefficients for the secondary component. In this case we had to lower the temperature used to evaluate the limb darkening coefficients only moderately to $T_{LD,1} = 4480$ K (scaling factor $a_2 = 4480 K / 5120 K \approx 0.88$) and the improvement in the obtained $\chi^2$ was much smaller -- see Fig.~\ref{fig:t2chi2p}. In fact the solution obtained for the LD coefficients corresponding to the effective star temperature $T_{2}=5120$ K was only a little more than 1-$\sigma$ inferior to the best one.

\subsection{Projection factor}
\label{sub:factor}
During the last decade there has been a substantial discussion about the proper projection factor ($p$-factor) to apply to observed Cepheid radial velocities to determine pulsational velocities. The issue came up when Gieren et al. (2005) tried to determine direct distances to Magellanic Cloud Cepheids by applying the near-infrared surface-brightness method to LMC Cepheids and found a non-physical period dependency of the derived distances. To correct for this they inferred, as the most likely explanation, a stronger variation of the $p$-factor than what had previously been assumed. This period effect has been observationally confirmed recently by Storm et al. (2011) who applied the surface-brightness method to a much larger sample of Cepheids. Based in large part on these new data Groenewegen (2013) and Ngeow et al. (2012) confirmed the stronger period dependence of the $p$-factor. Recent theoretical studies (e.g. Nardetto et al. 2009, Neilson et al. 2012) however do not predict that the $p$-factor should have a strong period dependence and they found significantly smaller values of the $p$-factor for short period Cepheids than inferred by the surface-brightness method studies. CEP-0227 provides a unique opportunity to directly measure the $p$-factor for a short period Cepheid. 

In our case the pulsations of the Cepheid star alter the shape of the light curve not only because its flux changes over the pulsation period but also because its radius does. This is manifested during the eclipses as the beginning and the end of the eclipse may be shifted in time and the visible area of the eclipsed star disk depends on the phase of the pulsating component. For any given moment of the eclipse this area is a function of the stars radii and the orbital inclination. As we know the area function from the light curve solution we can calculate directly the Cepheid radius and trace its changes for the given constant orbital parameters. Because the amplitude of the radius change scales with the projection factor (i.e. the larger the $p$-factor the more profound the radius change), measuring those changes we can directly constrain its value.
A conversion from the relative radii (used in the light curve analysis) to the absolute radii (used in the derivation of the $p$-factor) is done using the orbital solution previously obtained from the analysis of the radial velocities.

\begin{figure}
\begin{center}
  \resizebox{\linewidth}{!}{\includegraphics{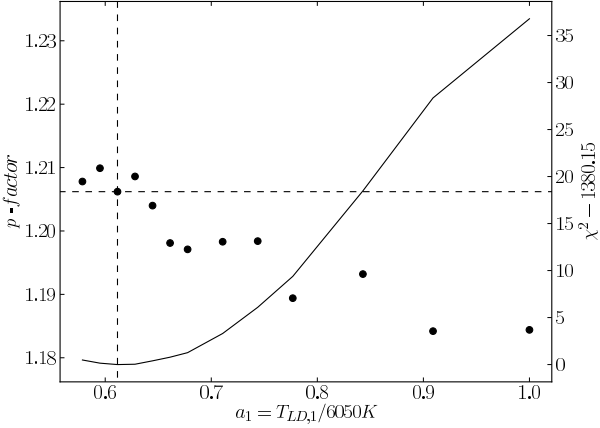}} \\
\caption{ The dependence of the projection factor (filled circles) and $\chi^2$ minimum value (solid line) on the temperature scaling factor $T_{LD,1}/6050$ for the pulsating component. The $p$-factor and the temperature scaling factor values for the best fit are marked with dashed lines. The y-axis span for the $p$-factor roughly corresponds to its estimated error (0.03).
\label{fig:t1chi2p}}
\end{center}
\end{figure}

\begin{figure}
\begin{center}
  \resizebox{\linewidth}{!}{\includegraphics{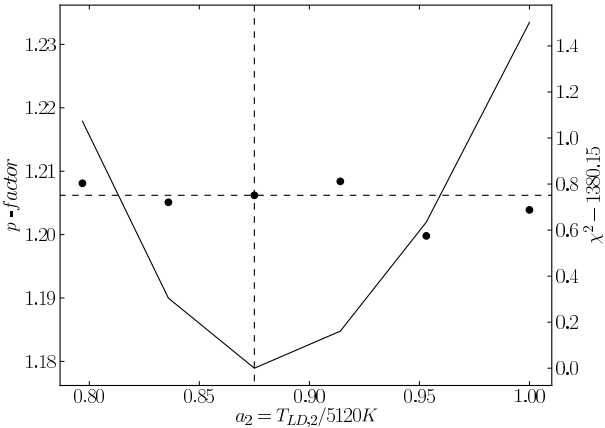}} \\
\caption{ Same plot as in Fig.~\ref{fig:t1chi2p} but for the temperature scaling factor $T_{LD,2}/5120$ for the companion star.
\label{fig:t2chi2p}}
\end{center}
\end{figure}

\begin{figure}
\begin{center}
  \resizebox{\linewidth}{!}{\includegraphics{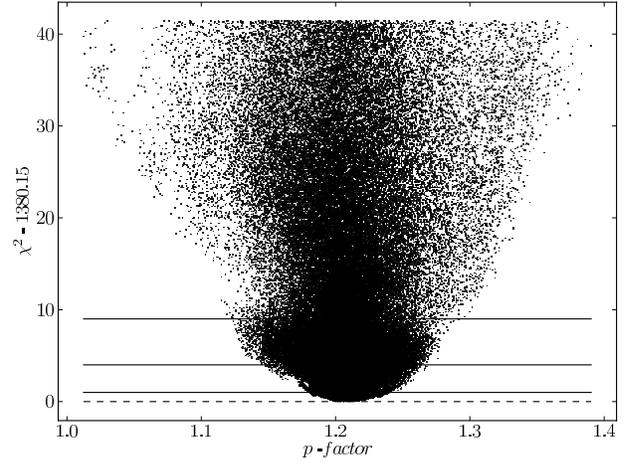}} \\
\caption{ A plot of $p$-factor in a $\chi^2$ plane obtained with the Monte Carlo simulation, 185,000 models are shown. The 1-, 2- and 3-$\sigma$ values are marked with solid horizontal lines and the zero level with a dashed one. Each point represents one calculated model. The lowest $\chi^2$ value was obtained for $p=1.206$, while central value is $ p=1.208$. The estimated 1-$sigma$ error is about $0.028$.
\label{fig:pfac}}
\end{center}
\end{figure}

The value of the projection factor which fits best our data is $p=1.206 \pm 0.028$, see Fig. \ref{fig:pfac}, and the uncertainty quoted is the statistical error estimated from the Monte Carlo simulations. This value is different from the value predicted by empirical calibration of Storm el al. (2011) for a Cepheid star with a period of 3.8 days which equals to $p=1.442$. We have tested the $p$-factor of 1.442 but the $\chi^2$ value was higher by more than 80, i.e. solution was about $9\sigma$ away from the best solution. 
However, such a low value of the projection factor is in good agreement with the theoretical calibrations by Nardetto at al. (2009) and Neilson et al. (2012). The mean wavelength of the spectral region we have used to derive radial velocities roughly corresponds to the effective wavelength of the Johnson V band. Thus, both calibrations for our Cepheid predict $p=1.26 \pm 0.03$ and $p=1.23 \pm 0.02$, respectively.  

It is important to note, that the $p$-factor value does not depend much on the used limb darkening coefficients -- see Fig. \ref{fig:t1chi2p} and \ref{fig:t2chi2p}. All the $p$-factors found for different LD coefficients sets are located within the range of 1.18-1.22, inside the 1-$\sigma$ border.  Because of this weak dependence we use the value and errors derived for the best set of LD coefficients as the final values. Another important thing is the complete independence of our approach to any assumptions on distance to OGLE-LMC-CEP-0227. In fact our photometric analysis is almost entirely done using only relative radii of the stars, which do not scale with distance. Also a conversion from the radial velocities to the pulsational ones is distance independent. 

To estimate a systematic uncertainty we compared all the determinations of the projection factor within all sorts of the investigated models (including those with different limb darkening coefficients, the third light neglected, etc.). This tells us how the determined value of the $p$-factor is sensitive to different model assumptions. In all cases the resulting $p$-factor lies within a range of 1.17 to 1.25. Thus, we assumed the systematic error of 0.04.

\subsection{Third light}
\label{sub:third}
The presence of the third light was investigated in our analysis. We allowed for its independent presence in each of the photometric bands. In the beginning the most suspicious was the {\it Spitzer} $3.6\mu$m band because some Galactic Cepheids were reported to have near-infrared excess (e.g. Kervella et al. 2006, M{\'e}rand et al. 2007) which is usually understood as a result of on-going mass loss. The solutions found, however, were consistent with no third light contribution in the {\it Spitzer} band and the V-band as well (being of the order of $0.1\%$). It turned out however, that some significant third light was present in the $I$-band light curve ($l_3=0.015$, i.e. $1.5\%$ of the total flux).  The detection of the third light only in the $I$-band is a bit surprising. It may indicate a presence of an unaccounted faint red blend in the OGLE photometry or some minor problems with the absolute calibration of the OGLE or {\it Spitzer} photometry. In fact Udalski et al. (2008) reported that the uncertainty of the absolute calibration of the OGLE photometry can reach 0.02 mag.

Taking the I-band third light into account results in a considerably smaller $\chi^2$ value with the detection on about 6-$\sigma$ level. A significant (more than 3-$\sigma$) difference in the obtained parameters between the models with and without the third light was found only in the case of the V-band surface brightness ratio. For the inclination, the Cepheid radius and $3.6\mu$m-band brightness ratio the difference is between 2 and 3 $\sigma$, and for the rest of the parameters the results are very consistent between solutions.

\subsection{Photometric parameters}
\label{sub:photo}
\begin{figure}
\begin{center}
  \resizebox{\linewidth}{!}{\includegraphics{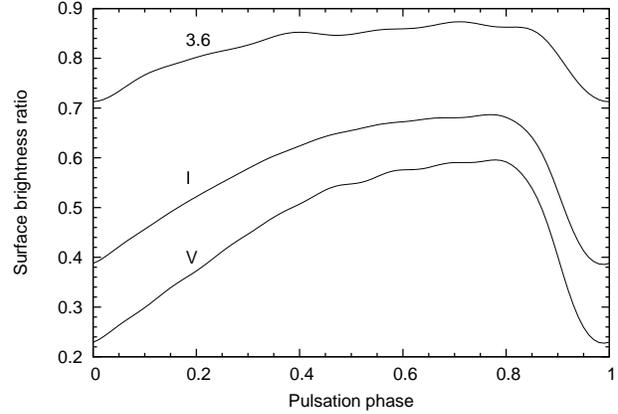}} \\
\caption{ Surface brightness ratio of the components $j_{21}$ as a function of the pulsation phase for three bands, from up to down: Spitzer 3.6$\mu$m, $I_C$, $V$.   
\label{fig:surf}}
\end{center}
\end{figure}

\begin{figure*}
\begin{center}
  \resizebox{\linewidth}{!}{\includegraphics{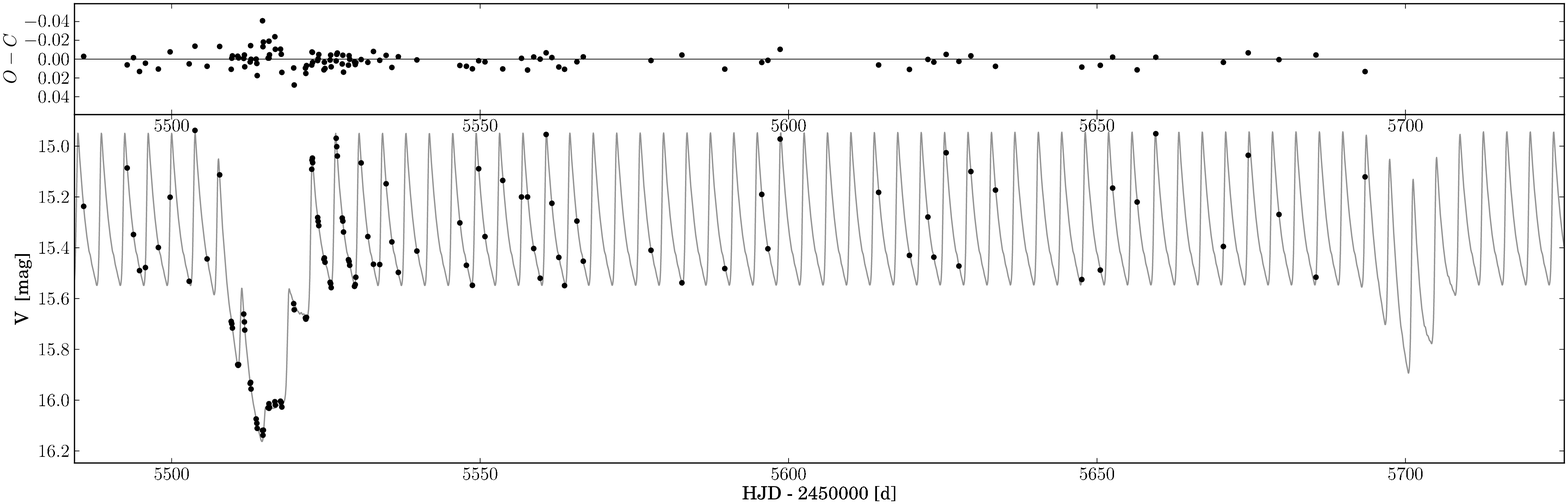}} \\
  \resizebox{\linewidth}{!}{\includegraphics{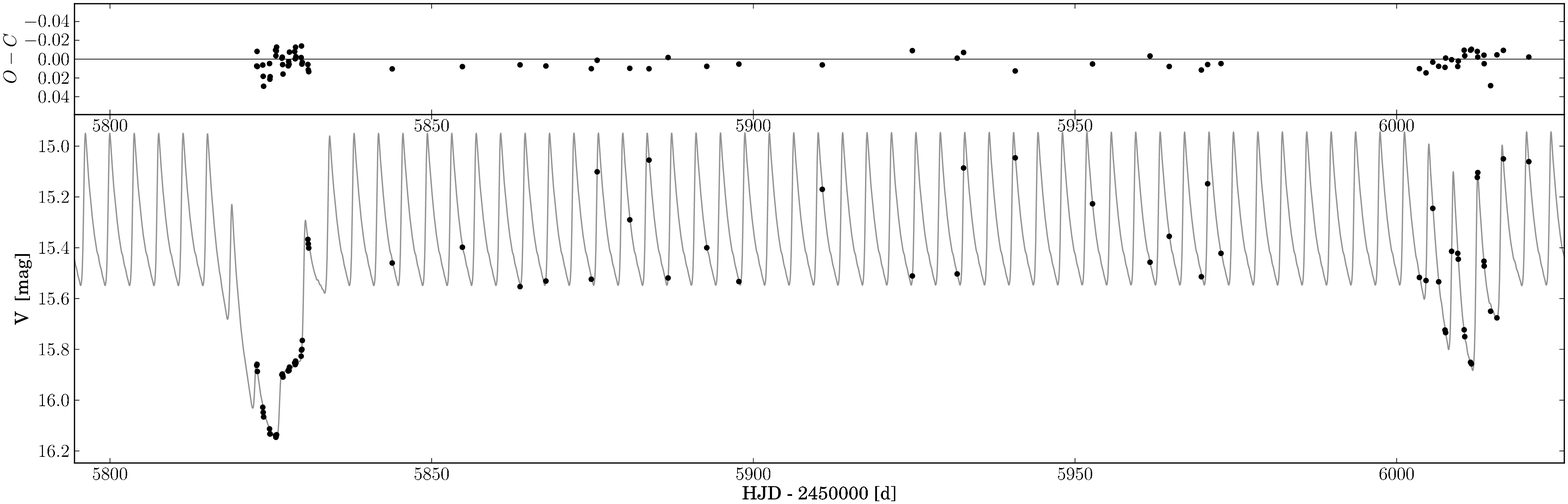}} \\
\caption{V-band model for selected eclipses. Observations are marked by small black circles.
\label{fig:vmodel}}
\end{center}
\end{figure*}

\begin{figure*}
\begin{center}
  \resizebox{\linewidth}{!}{\includegraphics{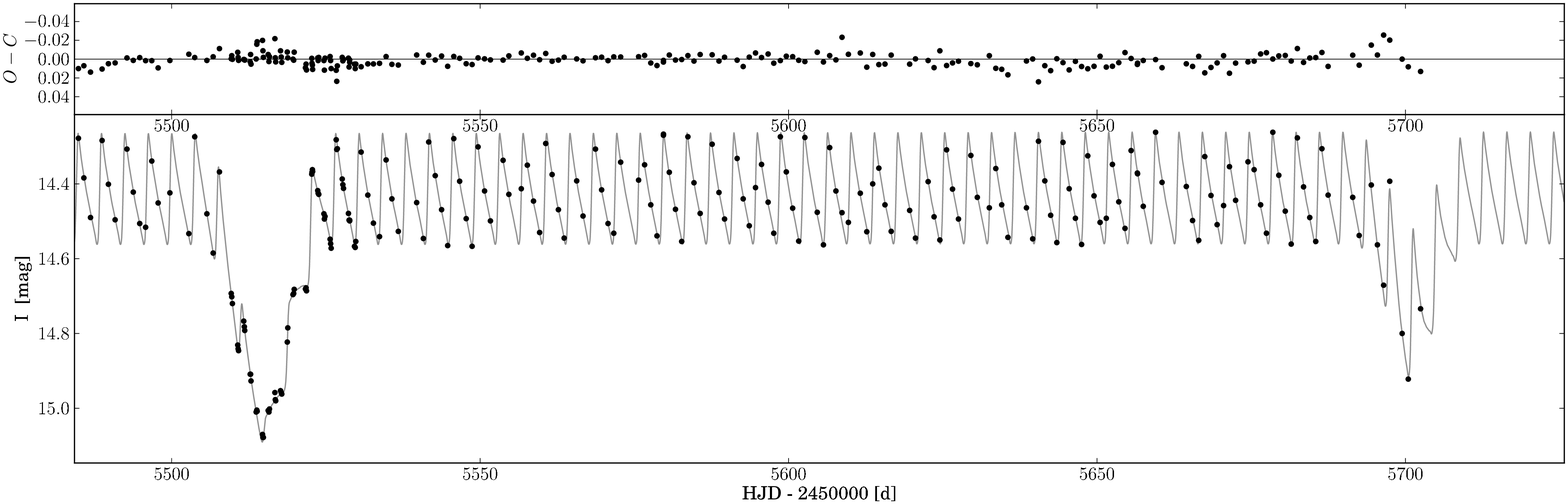}} \\
  \resizebox{\linewidth}{!}{\includegraphics{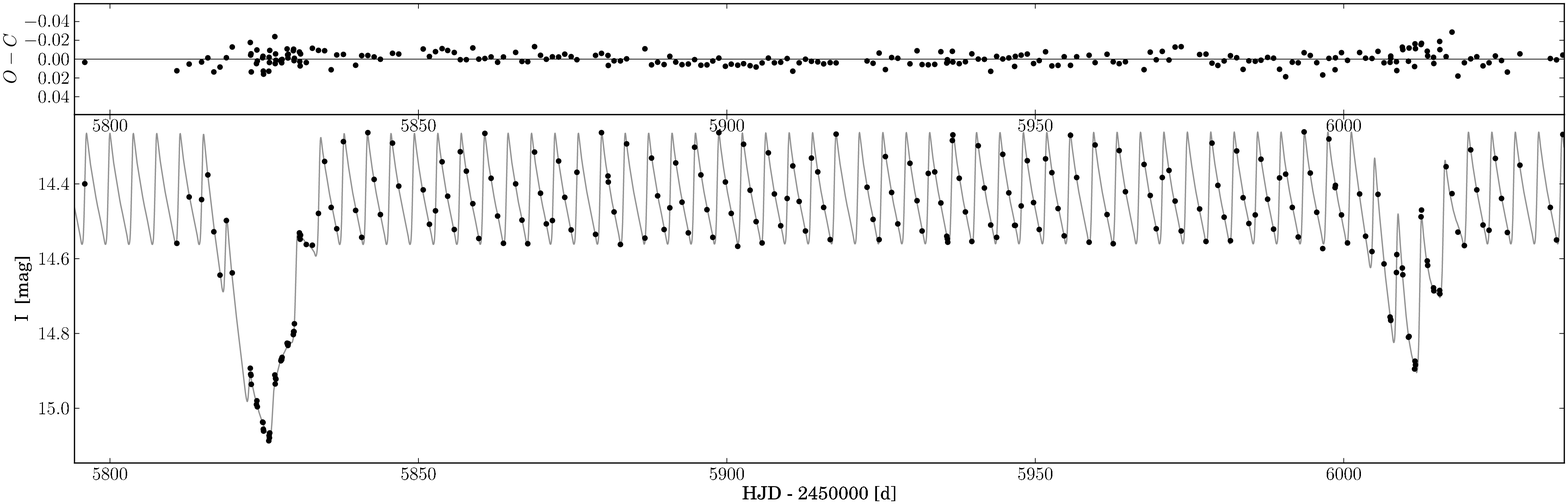}} \\
\caption{I-band model of selected eclipses.
\label{fig:imodel}}
\end{center}
\end{figure*}

\begin{figure*}
\begin{center}
  \resizebox{\linewidth}{!}{\includegraphics{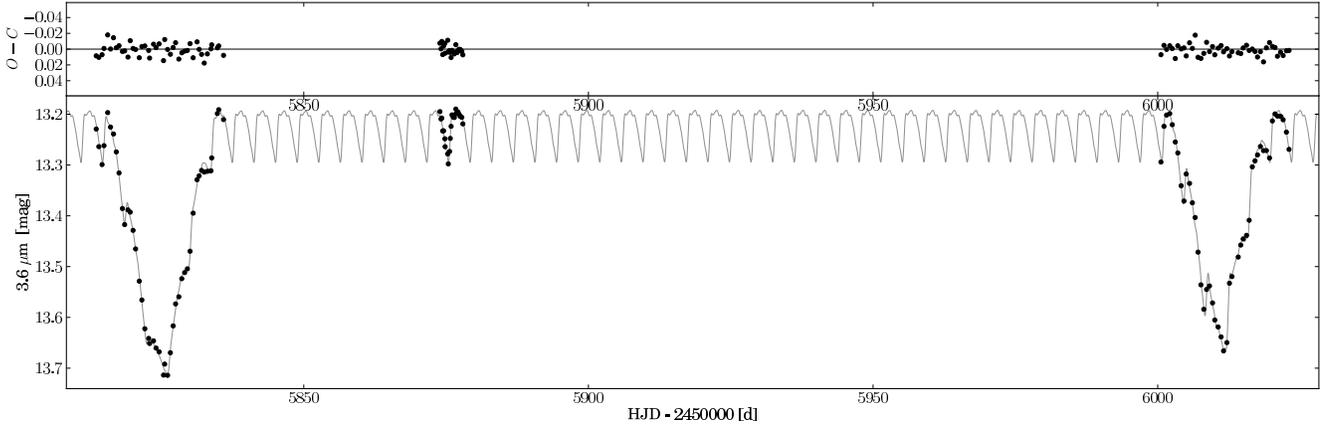}} \\
\caption{Spitzer 3.6 $\mu$m-band model.
\label{fig:lmodel}}
\end{center}
\end{figure*}

\begin{figure}
\begin{center}
  \resizebox{\linewidth}{!}{\includegraphics{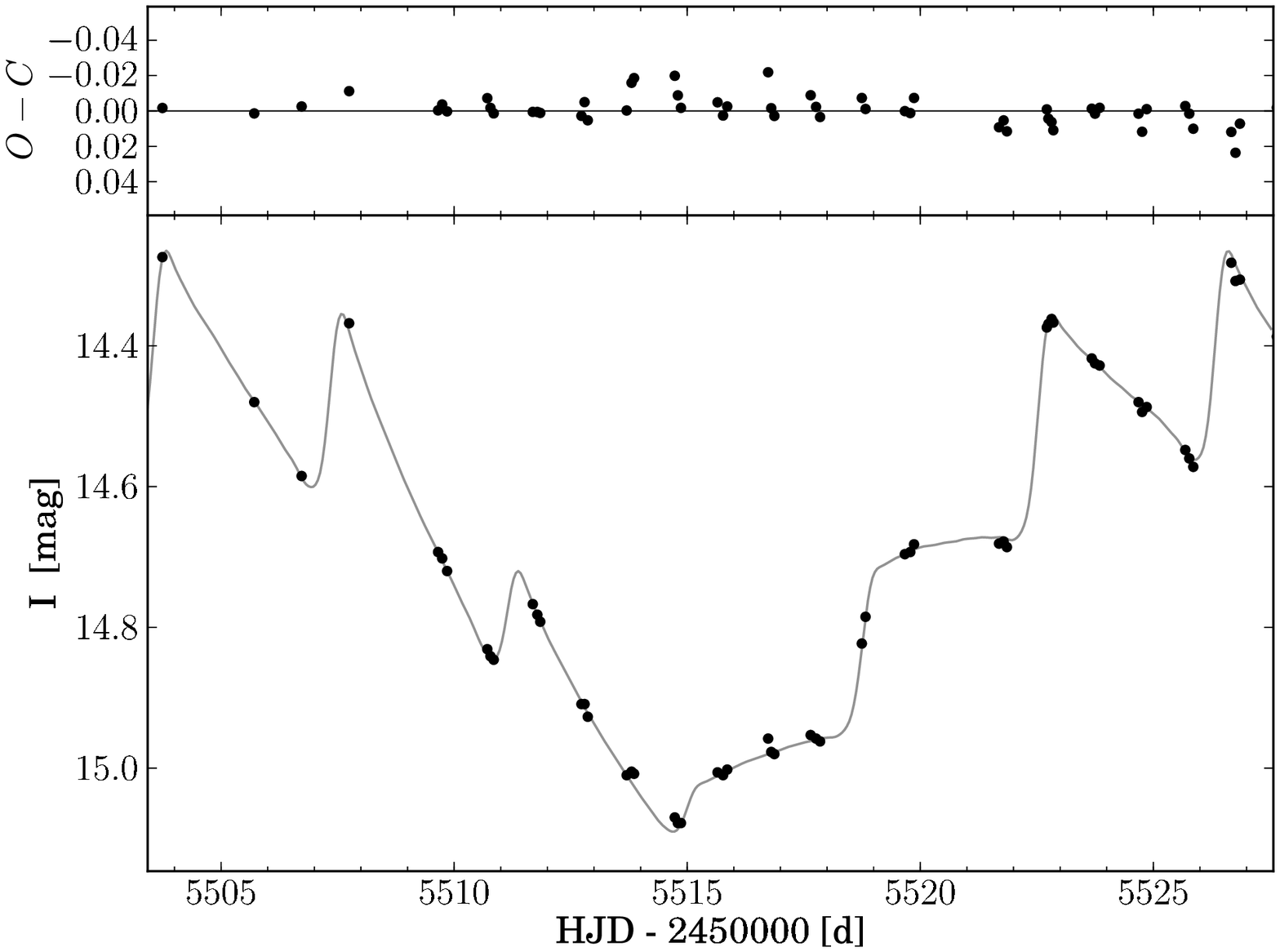}} \\
  \resizebox{\linewidth}{!}{\includegraphics{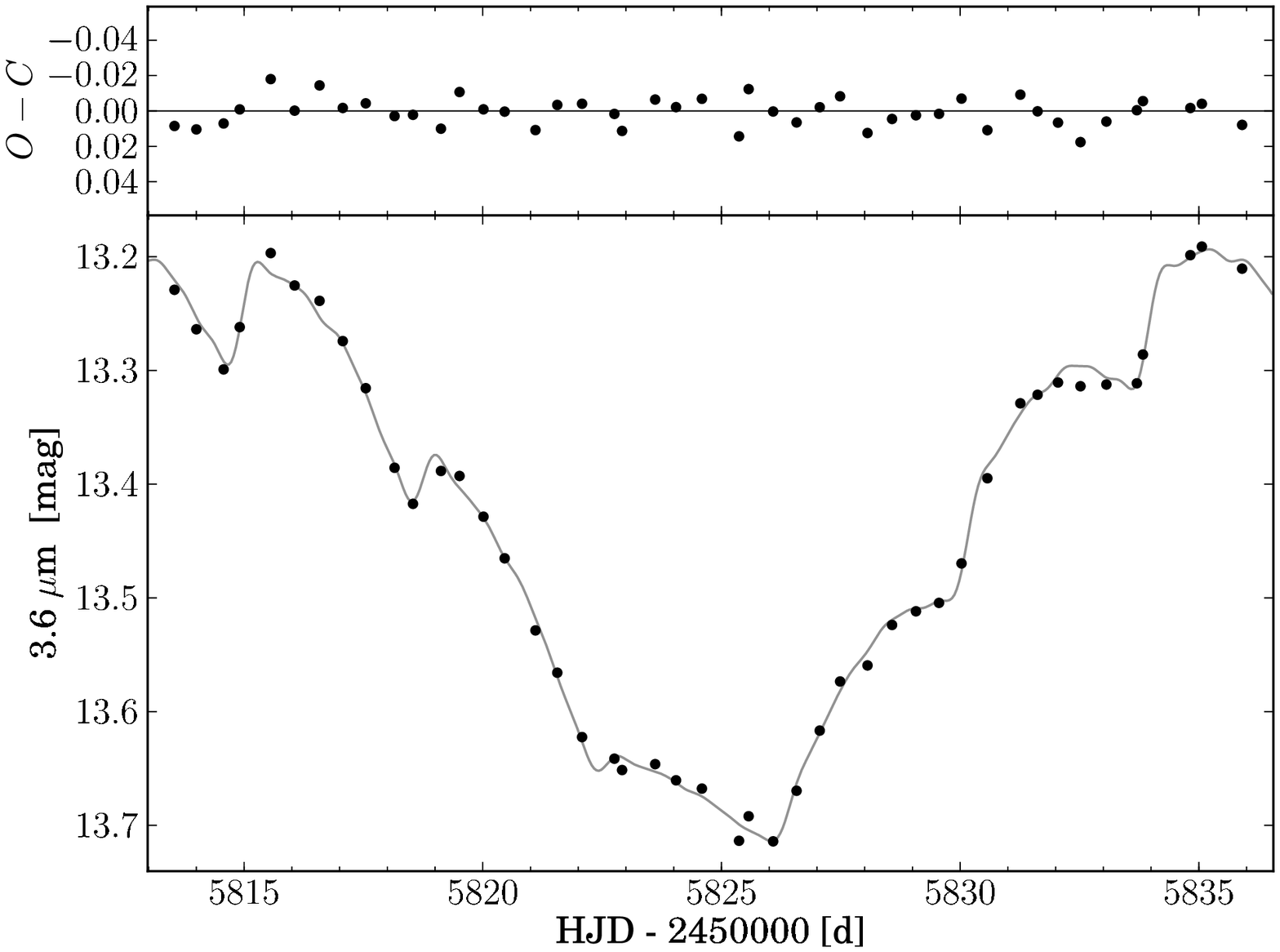}} \\
\caption{Spitzer 3.6 $\mu$m and I-band model shown for two different primary eclipses.
\label{fig:modelzoom}}
\end{center}
\end{figure}

The photometric parameters for our best solution, with the third light in the $I$-band taken into account, are summarized in Table~\ref{tab:photpar}. The light curve solution for all three photometric bands is presented in Fig.~\ref{fig:vmodel}--\ref{fig:modelzoom}. The model usually predicts well the brightness of the system during eclipses, however some small systematic residua are still present. The amplitude of the pulsations during the primary eclipse is smaller because a significant part of the Cepheid disk is covered at this stage and thus, relatively more light comes from the constant component. During the secondary (shallower) eclipse the Cepheid transits across the companion disk and the observed amplitude of the pulsations grows larger. In the near-infrared the pulsations become much less prominent and so they affect the shape of the eclipses less. Also the surface brightness ratio of the components $j_{21}$ changes considerably from the optical to near-infrared. Fig.~\ref{fig:surf} presents the dependency of $j_{21}$ on the pulsation phase and the photometric band for our best model.

\begin{table}
\caption{Photometric parameters of CEP-0227 from the Monte Carlo simulations. Values marked with $^{\textstyle a}$ correspond to a pulsation phase 0.0, $T_0$ is $HJD - 2450000$~d. Limb darkening coefficients for the logarithmic law are presented -- all of them were adjusted simultaneously using a single parameter (see text for details). $L_{21}$ is the light ratio of the components in every photometric band.} 
\begin{tabular}{@{}l|c|c|l@{}} \hline
Parameter & Mean value & Best fitted value & Error \\\hline 
Adjusted & & & \\
$P_{obs} (d) $ & - & 309.6690 & 0.0017  \\
$T_0$ (d) & - & 4895.908 & 0.005\\
$r_1$ & 0.08957 & 0.08532$^{\,\textstyle a}$ & 0.00025 \\
$r_2$ &  - & 0.11503 & 0.00025 \\
$j_{21}(V)$& 0.4566 & 0.2296$^{\,\textstyle a}$& 0.0015 \\
$j_{21}(I_C)$& 0.5791 & 0.3881$^{\,\textstyle a}$& 0.0015 \\ 
$j_{21}(3.6)$& 0.8206 & 0.7146$^{\,\textstyle  a}$ & 0.0045 \\ 
$i$ ($^\circ$) & - & 86.833 & 0.016 \\
$e$ & - & 0.1659 & 0.0006 \\
$\omega$ ($^\circ$)& - & 342.0 & 0.6\\
$p$-factor & - & 1.206 & 0.030\\
$l_{3,V}$  & -& 0.000 & 0.002 \\
$l_{3,I}$  & 0.018 & 0.015$^{\,\textstyle  a}$ & 0.002 \\ 
$l_{3,3.6}$ & -& 0.000 & 0.002 \\
$u_{1,V}$& &\multicolumn{2}{l}{0.805 \, $-0.166$ } \\
$u_{1,I}$ & &\multicolumn{2}{l}{0.648 \,\,\,\, 0.129}  \\
$u_{1,3.6}$ & &\multicolumn{2}{l}{0.375 \,\,\,\, 0.218}  \\
Derived quantities& & & \\
$L_{21}(V)$ & 0.7504 & 0.4174$^{\,\textstyle a}$ & \\ 
$L_{21}(I_C)$ & 0.9539& 0.7054$^{\,\textstyle a}$  &\\ 
$L_{21}(3.6)$ & 1.357 & 1.299$^{\,\textstyle a}$ & \\\hline
\label{tab:photpar}
\end{tabular}
\end{table}

\begin{figure*}
\begin{center}
  \resizebox{0.49\linewidth}{!}{\includegraphics{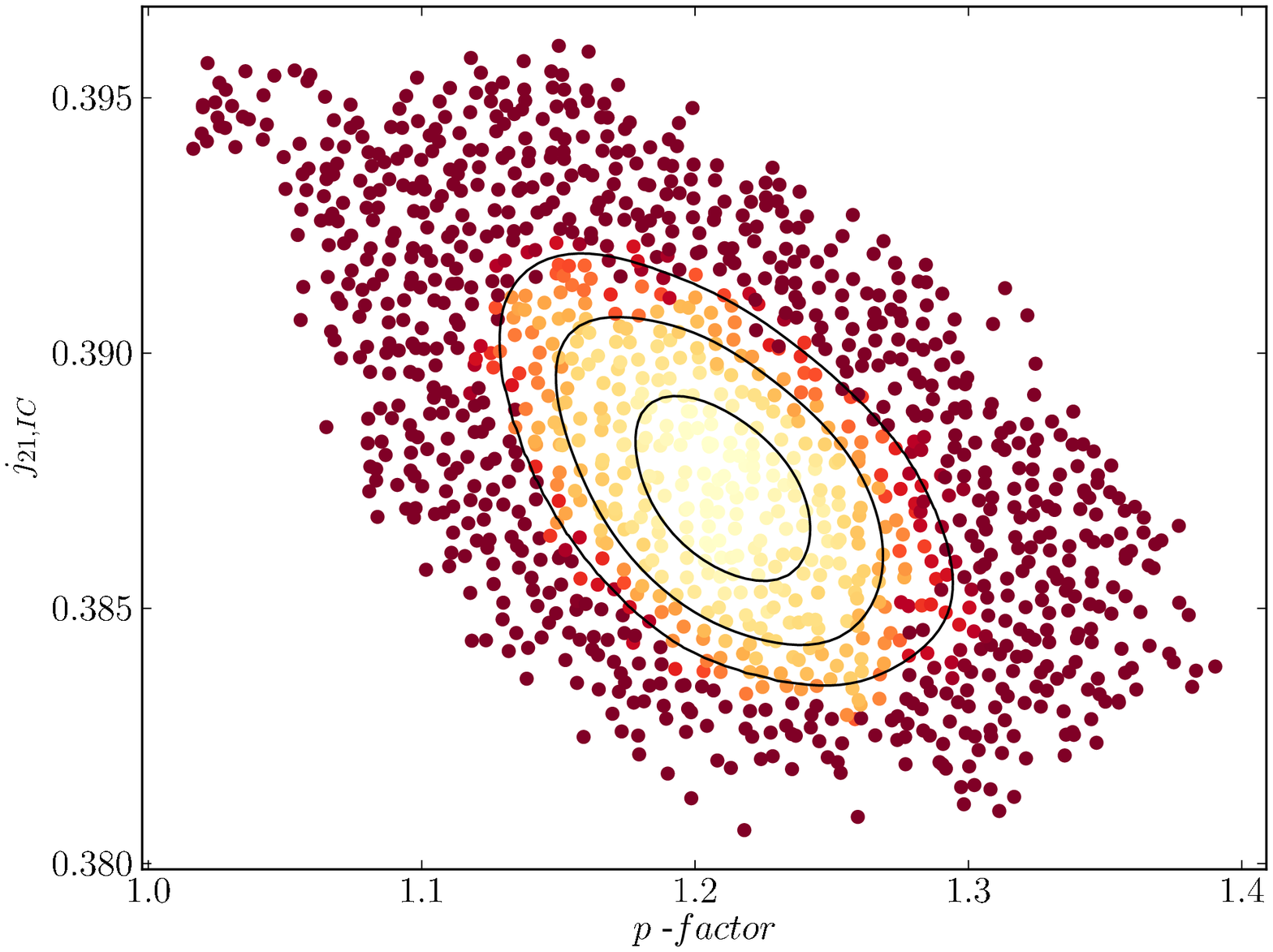}} 
  \resizebox{0.49\linewidth}{!}{\includegraphics{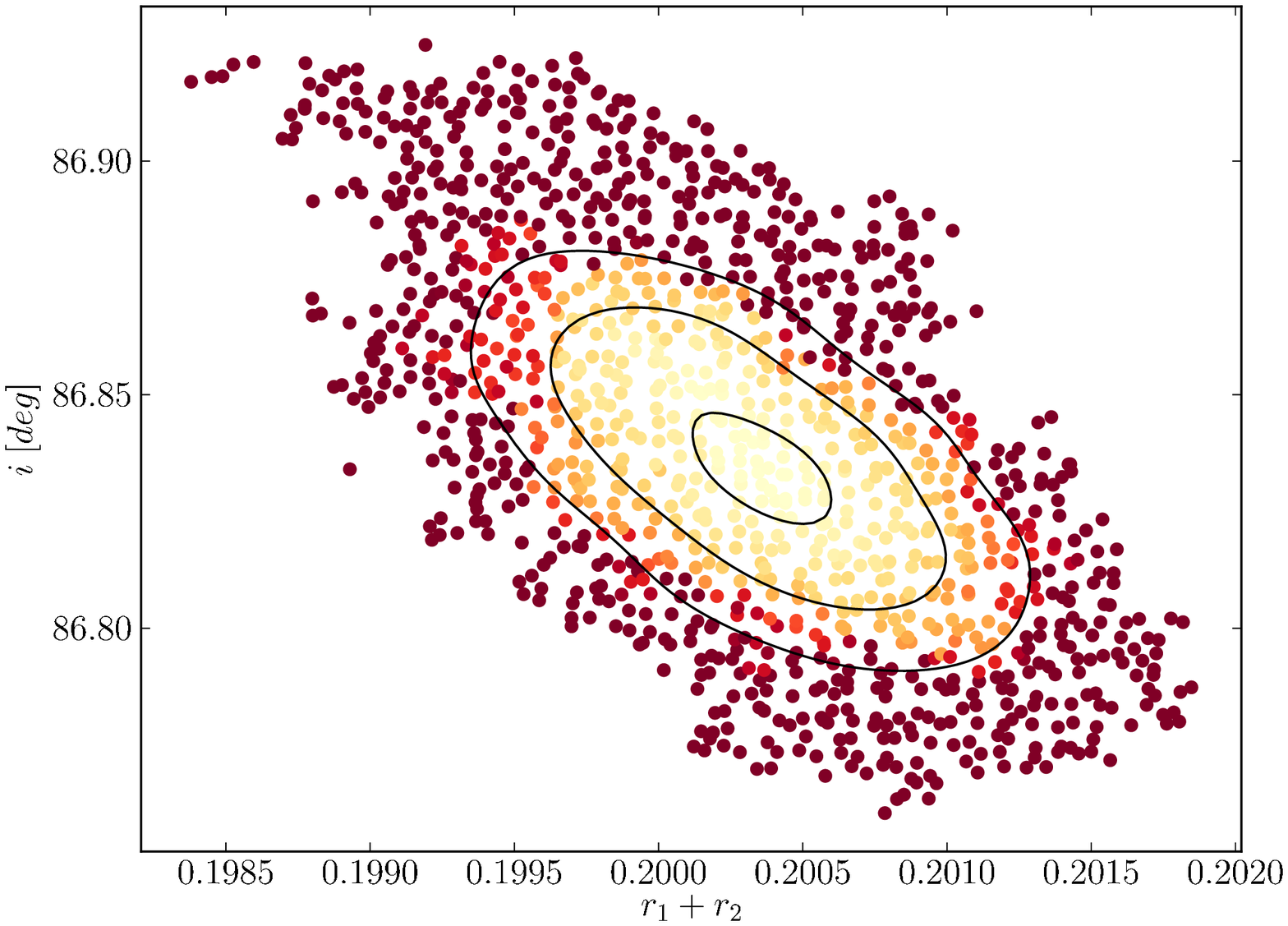}} 
  \resizebox{0.49\linewidth}{!}{\includegraphics{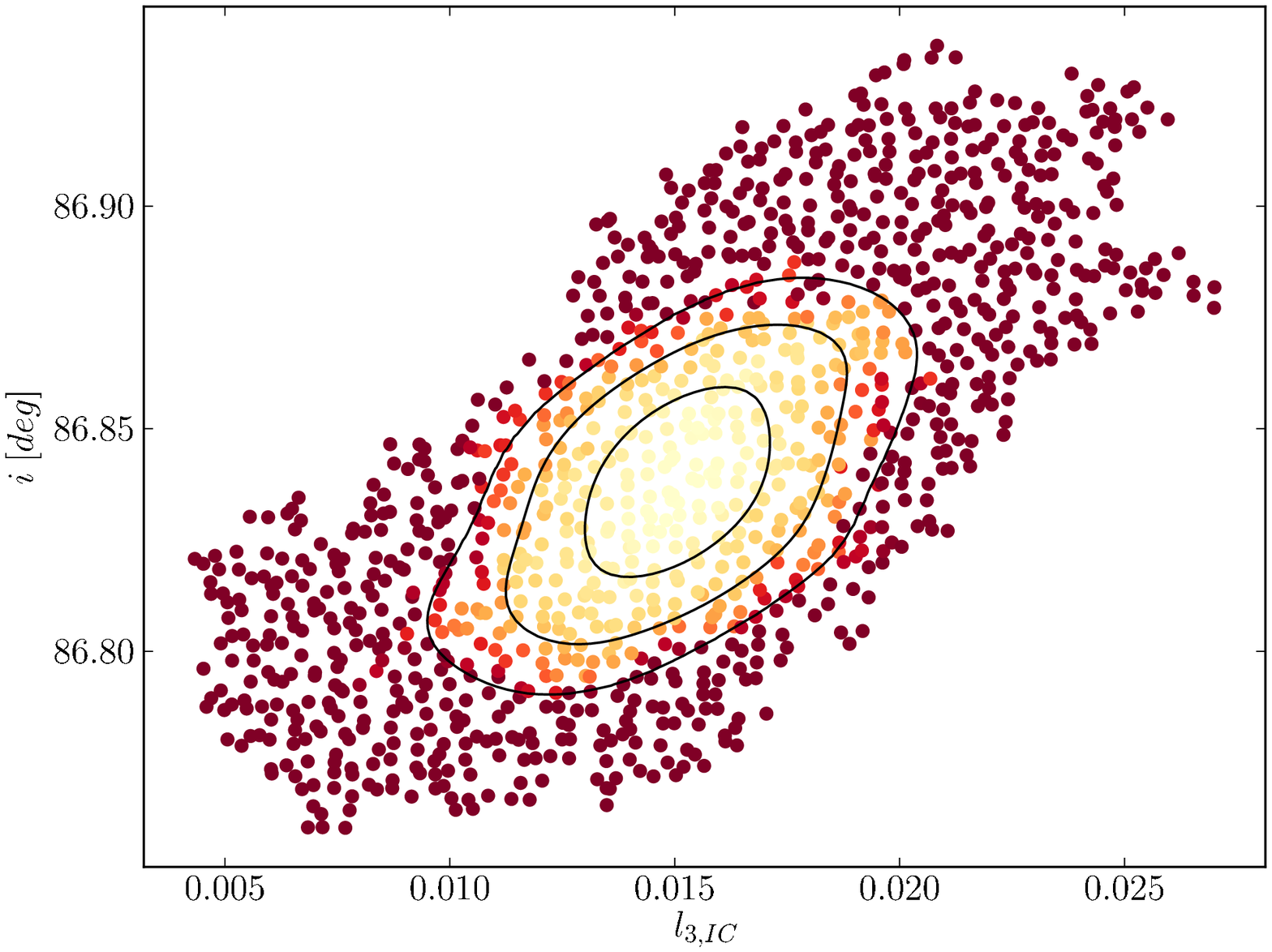}} 
  \resizebox{0.49\linewidth}{!}{\includegraphics{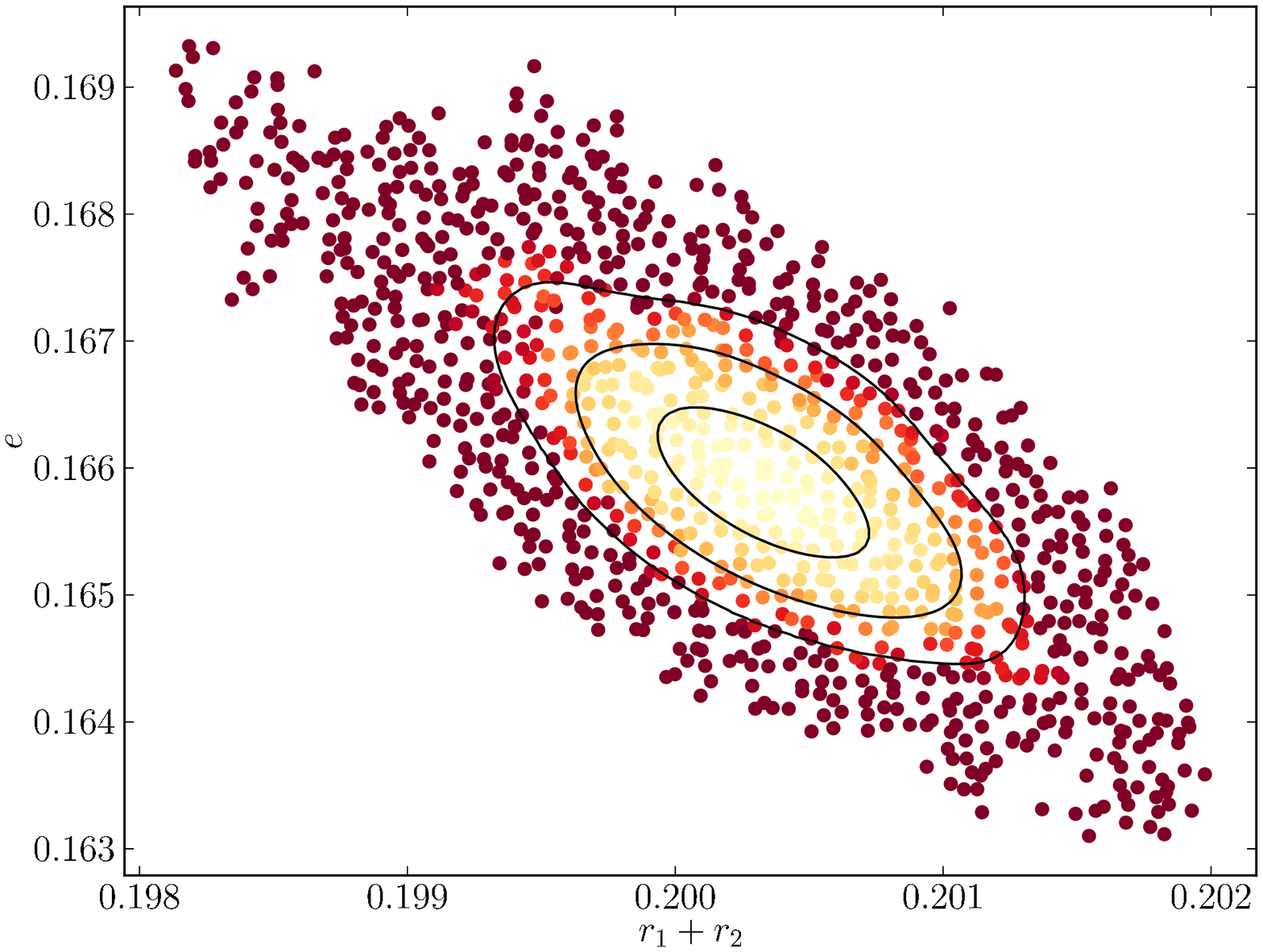}} 
\caption{Correlations between the $p$-factor and the I-band surface brightness ratio, the inclination and the sum of the star fractional radii, the inclination and the third light in the I-band, the eccentricity and the sum of the star radii. The best models for given pair of the parameters are shown and the $\chi^2$ values are coded with color (higher values are darker). Solid lines represent 1-, 2- and 3-$\sigma$ levels for the two-parameter error estimation.
\label{fig:corr}}
\end{center}
\end{figure*}

Most of the parameters fitted in our approach are independent and do not exhibit any significant correlation, though some do. We were concerned how the projection factor correlates with the other photometric parameters but in this case we only detected a weak correlation with the surface brightness ratios. In Fig.~\ref{fig:corr} the correlation with the $I$-band surface brightness ratio is presented, which is the main source of the statistical uncertainty on the determined $p$-factor value. The strongest correlation among the parameters in our solution was found between the orbital plane inclination $i$ and the sum of the radii $r_{1}+r_{2}$ (same figure) and it is the prime error source of the absolute radii uncertainty. The aforementioned correlation between the inclination and the third light, as well as between the eccentricity and the sum of the fractional radii are also presented.

\subsection{Absolute dimensions}

\begin{table}
\caption{Physical properties of CEP-0227. The spectral type, radius, gravity ($\log g$), temperature, luminosity ($\log L$) and the observed magnitudes are mean values over the pulsation period. The orbital period is a rest frame value.}
\begin{tabular}{@{}l|c|c@{}} \hline
Parameter & Primary (Cepheid) & Secondary \\\hline
spectral type & F7 Ib & G4 II \\
mass ($M_\odot$) & 4.165 $\pm$ 0.032 & 4.134 $\pm$ 0.037 \\ 
radius ($R_\odot$) & 34.92 $\pm$ 0.34  & 44.85 $\pm$ 0.29 \\ 
$\log g$ (cgs) & 1.971 $\pm$ 0.011 & 1.751 $\pm$ 0.010  \\ 
temperature (K) & 6050 $\pm$ 160 & 5120 $\pm$ 130 \\ 
$\log L$ ($L_\odot$) & 3.158 $\pm$ 0.049 & 3.097 $\pm$ 0.047 \\
$V$ (mag) &15.932 & 16.244\\
$I$ (mag) &15.178 & 15.229\\
$K$ (mag) &14.221 & 13.903\\
$v\sin i$ (km/s) & - & 11.1 $\pm$ 1.2 \\ 
orbital period (days) & \multicolumn{2}{c}{309.404 $\pm$ 0.002 } \\
semimajor axis ($R_\odot$) & \multicolumn{2}{c}{389.86 $\pm$ 0.77} \\\hline
\label{tab:abs}
\end{tabular}
\end{table}
Table~\ref{tab:abs} presents the physical parameters of both components and some orbital parameters as well. The spectral type is estimated from the effective temperature scale given in Table~1 of Alonso et al. (1999). The luminosity class is taken from Ginestet et al. (2000). The surface temperatures of the components were calculated according to the dereddened ($V\!-\!K$) colors (Gieren et al. in preparation). The effective temperature of the primary was independently derived by Marconi et al. (2013) as $T_1 = 6100$ K and it is in good agreement with our estimate. 

The total error of the absolute radii determination contains statistical uncertainties from the relative radii and the semi-major axis determination. Additionally we add some systematic uncertainty to the budget error which comes from the presence of the small systematic residua still existing in our photometric solution. During eclipses the magnitude of these residua reaches 0.01 mag which translates into $0.9\%$ uncertainty of the flux and $0.45\%$ uncertainty of the radii. One must note however that the similar systematic residua are also present {\em outside} the eclipses and as such may be attributed to some defects of the photometry and not to the model itself.
The final error was eventually derived as a sum of all the partial errors in quadrature. The Cepheid is the largest at the pulsation phase $\phi=0.40$ reaching $R_{max}=36.44 \, R_\odot$ and the smallest at the pulsation phase $\phi=0.92$ shrinking to $R_{min}=32.43 \, R_\odot$.   

\begin{figure}
\begin{center}
  \resizebox{\linewidth}{!}{\includegraphics{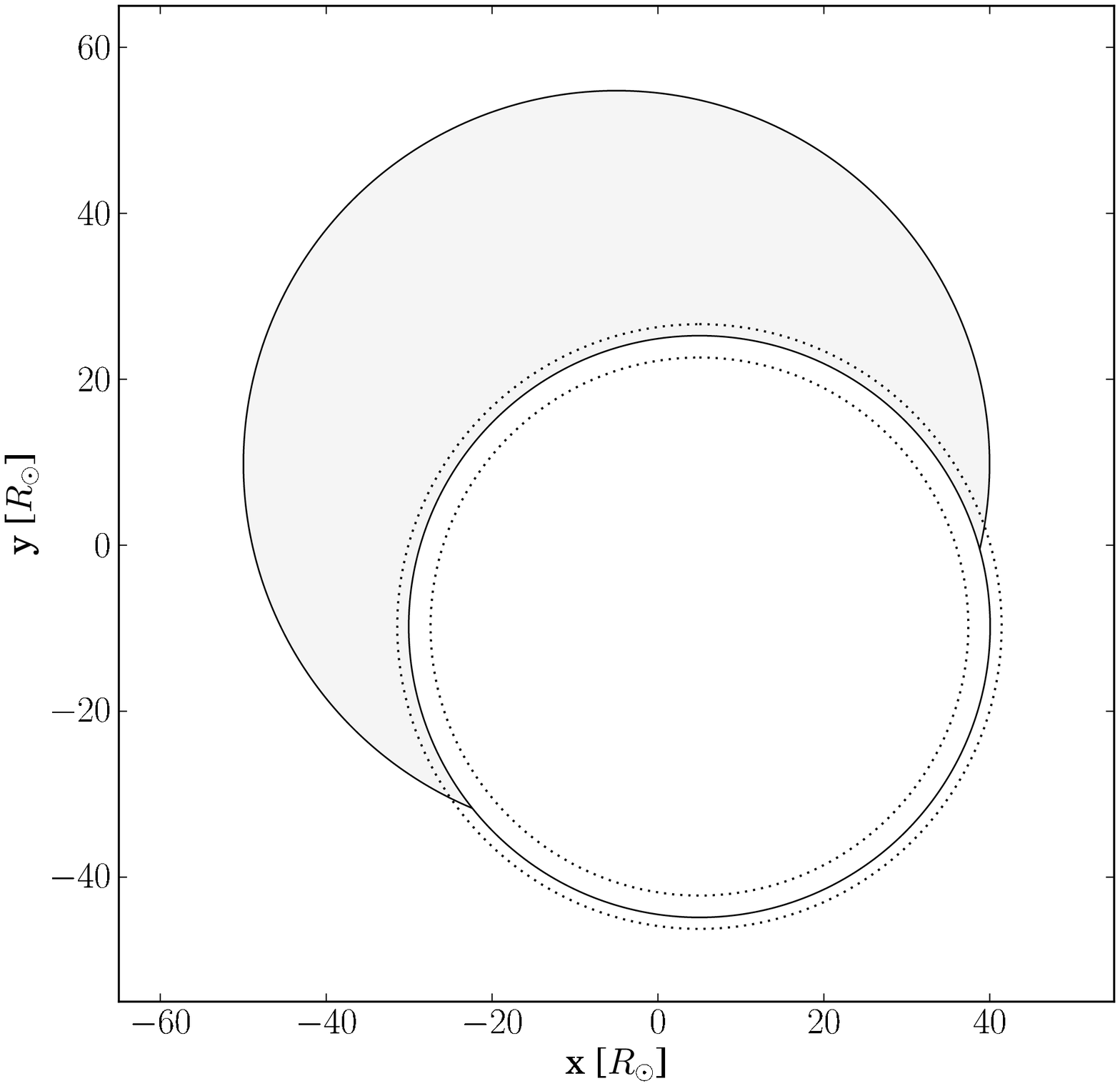}} \\
\caption{System configuration close to the secondary mid-eclipse. The~Cepheid is passing in front of the red giant companion. Star edge line width represents 1-$\sigma$ formal error in the determination of the radii. The Cepheid radius is a mean value over time and dotted lines represent its minimum and maximum radii. Changes in the radial amplitude of the star that correspond to the $p$-factor error approximately equal the radius error. The distance between the stars at this phase is about 355 $R_\odot$.
\label{fig:eclipse}}
\end{center}
\end{figure}

The rotation is derived from the Broadening Function calculated for all the spectra where the components are well-separated. The profile is wide and the instrumental broadening seems to be of secondary importance. We assumed that the rotation axis is perpendicular to the orbital plane. The derived value of the projected rotational velocity of the secondary $v_2\sin{i}=11.1$ km/s is consistent with its pseudosynchronous rotation of 10.4 km/s. The rotational velocity of the primary is strongly affected by the atmospheric turbulence originating from the pulsations, so we do not determine this parameter here.

It is worth to mention that the new radius, mass and luminosity estimates do agree within one sigma with the recent pulsation and evolutionary prescriptions.

\section{Conclusions}
\label{sect:concl}
The presented method proofed to be a good tool for the analysis of eclipsing binaries with radially pulsating components. It allows for a consistent treatment of the photometric and spectroscopic data -- calculating the pulsating component radius change we make use of both of them. As a result, very precise measurements of the physical parameters of a Cepheid variable and its companion were obtained. We fully confirmed the findings of Pietrzy{\'n}ski et al. (2010), especially the reported mass and radius values of the classical Cepheid OGLE-LMC-CEP-0227. Our masses for both components and the radius of the secondary are well within 1-$\sigma$ error bars given by Pietrzy{\'n}ski et al. (2010). A slight difference occurs for the Cepheid mean radius which is about $1.7\sigma$ larger in our solution. We do not think that to be significant because we have analyzed here a much larger set of observations and the previous analysis was based on the approximate removal of the pulsations from the light curve. Our mean radius is in perfect agreement with the Cepheid period - radius relation of Gieren et al. (1999) and marginally consistent with Groenewegen (2013) calibration. 

The present analysis of the {\it Spitzer} data excludes the possibility of the additional third light in the near-infrared larger than $\sim0.2\%$. Because the level of the third light detected in the I-band is also low we conclude that there is no significant K-band excess in this system as well.

The observed disk of the Cepheid surface seems to be heavily darkened, especially in the optical region where corresponding linear LD coefficient is $u_V\approx 0.9$. It is at odds with the limb darkening coefficient predicted for the static atmosphere at the temperature $T=6050$ K, the gravity $\log{g}=1.97$ and the metallicity $[$Fe/H$]=-0.5$, namely $u_V=0.56$ (van Hamme 1993). Such a strong limb darkening may arise from the high degree of turbulence in the pulsating atmosphere of the Cepheid and from the presence of very deep and profound convective cells more typical for a late K-type giant.

According to our knowledge the method we have used for deriving the projection factor is the first one of this kind reported in the literature. It is also a second time, in a general case of short period Cepheids, that the individual value is  precisely determined after the interferometric measurements for $\delta$ Cep (M{\'e}rand et al. 2005). Our value of the projection factor $p=1.21 \pm 0.03$ is close to the $p$-factor determined by M{\'e}rand et al. (2005) $p=1.27 \pm 0.06$. Marconi et al. (2013) basing on the hydrodynamic models of the OGLE-LMC-CEP-0227 derived the $p$-factor $p=1.20 \pm 0.08$, in good agreement with our empirical determination. However, their models were fitted to the pulsation light curves of the Cepheid which were freed from the companion light contribution according to our photometric light curve solution. Thus, their value of the $p$-factor is not fully independent.

There are two substantial advantages of our method in comparison with the approach presented by M{\'e}rand et al. (2005) making it less prone to systematics. First, our projection factor is distance independent.  Second, only weak dependence on the limb darkening assumptions is present. In fact, limb darkening coefficients are fitted simultaneously, but independently to the $p$-factor. Let us emphasize here that in deriving the interferometric angular diameters one need to convert uniform disk diameters $\theta_{UD}$ into limb darkened ones $\theta_{LD}$. For $\delta$ Cep the conversion was done using the theoretical limb darkening tables for ordinary (non-pulsating) stars. However, in a view of the peculiar limb darkening we have found for the CEP-0227 such procedure may be called into question. 

Of course there are some other sources of possible systematics in our solution. First of all, the question if JKTEBOP can adequately represent the surfaces of giant stars. Comparison made with the Wilson-Devinney code (Graczyk et al. 2012), which is still the most elaborated program for the analysis of eclipsing binaries, suggests that for well detached binaries (as our CEP-0227) the solutions returned by both codes are very similar. If any systematics connected with the use of JKTEBOP exists, most probably it is shared by other computer tools for modeling eclipsing binaries. Some systematics may arise also from the assumptions of constant limb darkening and projection factor during the whole pulsation cycle. The validation of both assumption is currently under work and will be presented in another paper.

The application of the light travel time effect was important in the analysis. For our object it barely affected our derived parameters but the overall fit was significantly better removing some systematic residuals. For some other objects like OGLE-LMC-CEP-1812 we expect the effect to have an even higher impact on the solution.

In summary we conclude that the presented method allowed us not only to improve the precision of the determination of the intrinsic and structural parameter of the binary system and the pulsating component in particular, but also to measure some other characteristics like limb-darkening and the p-factor.  It has a great potential for the application to other binary systems with radially pulsating components.

\section*{Acknowledgements}
\label{sect:acknow}
We gratefully acknowledge financial support for this work from the Polish National Science Center grant MAESTRO 2012/06/A/ST9/00269 and the TEAM subsidy from the Foundation for Polish Science (FNP). Support from the BASAL Centro de Astrof{\'i}sica y Tecnolog{\'i}as Afines (CATA) PFB-06/2007 is also acknowledged. AG acknowledges support from FONDECYT grant 3130361. RS is supported from the Polish NSC grant UMO-2011/01/M/ST9/05914.

This work is based (in part) on observations made with the Spitzer Space Telescope, which is operated by the Jet Propulsion Laboratory,
California Institute of Technology under a contract with NASA. Support for this work was provided by NASA. The OGLE project has received funding from the European Research Council under the European Community's Seventh Framework Programme (FP7/2007-2013) / ERC grant agreement no. 246678 to AU.

We would like to thank the support staff at the ESO Paranal and La Silla observatory and at the Las Campanas Observatory for their help in obtaining the observations and the rest of the OGLE team for their contribution in acquiring the data for the object.

This research has made use of NASA's Astrophysics Data System Service.


\bsp
\label{lastpage}

\end{document}